\documentclass[aps,pre,amsmath,amssymb,notitlepage,reprint,10pt,longbibliography,superscriptaddress]{revtex4-2}
\usepackage{amsmath,epsfig,amssymb}
\usepackage{bbm}
\usepackage{times}
\usepackage{color}
\usepackage{amsthm}
\usepackage{color}
\usepackage[caption=false]{subfig}
\usepackage{multirow}
\usepackage[normalem]{ulem}
\usepackage{amsfonts,amssymb,dsfont}
\usepackage{graphicx}
\usepackage{soul}
\usepackage{hyperref}
\hypersetup{
    colorlinks=true,       
    linkcolor=red,          
  citecolor=magenta,        
    filecolor=magenta,      
    urlcolor=cyan,           
    runcolor=cyan
}
  
\usepackage[acronym,shortcuts]{glossaries}
\usepackage{tikz}
\usepackage{tikz-network}

\newcommand{\todo}[1]{}
\renewcommand{\todo}[1]{{\color{red} TODO: {#1}}}
\newcommand{\note}[1]{}
\renewcommand{\note}[1]{{\color{red} NOTE: {#1}}}
\newcommand{\question}[1]{}
\renewcommand{\question}[1]{{\color{red} QUESTION: {#1}}}

\newacronym{GSP}{GSP}{graph signal processing}
\newacronym{GFT}{GFT}{graph Fourier transform}
\newacronym{TDA}{TDA}{topological data analyis}
\newacronym{DTFT}{DTFT}{discrete time Fourier transform}
\newacronym{TV}{TV}{total variation}

\begin{document} 

\glsdisablehyper

%
\title{Analyzing Collective Motion Using Graph Fourier Analysis}
\author{Kevin Schultz}
\affiliation{Johns Hopkins University Applied Physics Laboratory, Laurel, MD 20723, USA}
\author{Marisel Villafa\~{n}e-Delgado}
\affiliation{Johns Hopkins University Applied Physics Laboratory, Laurel, MD 20723, USA}
\author{Elizabeth P. Reilly}
\affiliation{Johns Hopkins University Applied Physics Laboratory, Laurel, MD 20723, USA}
\author{Grace M. Hwang}
\affiliation{Johns Hopkins University Applied Physics Laboratory, Laurel, MD 20723, USA}
\affiliation{Kavli Neuroscience Discovery Institute, Johns Hopkins University, Baltimore, MD 21218, USA}
\author{Anshu Saksena}
\affiliation{Johns Hopkins University Applied Physics Laboratory, Laurel, MD 20723, USA}

\begin{abstract}

    Collective motion in animal groups, such as swarms of insects, flocks of birds, and schools of fish, are some of the most visually striking examples of emergent behavior. Empirical analysis of these behaviors in experiment or computational simulation primarily involves the use of ``swarm-averaged'' metrics or order parameters such as velocity alignment and angular momentum. Recently, tools from computational topology have been applied to the analysis of swarms to further understand and automate the detection of fundamentally different swarm structures evolving in space and time.  Here, we show how the field of graph signal processing can be used to fuse these two approaches by collectively analyzing swarm properties using graph Fourier harmonics that respect the topological structure of the swarm. This graph Fourier analysis reveals hidden structure in a number of common swarming states and forms the basis of a flexible analysis framework for collective motion.

\end{abstract}

\maketitle






\section{Introduction}

Collective motion  in animal groups \cite{boinski2000move,sumpter2010collective,vicsek2012collective}, such as swarms of insects \cite{seeley1979natural,schultz2008mechanism,kelley2013emergent}, flocks of birds \cite{ballerini2008empirical,cavagna2014bird}, and schools of fish \cite{herbert2011inferring,ioannou2012predatory}, are some of the most visually striking examples of emergent behavior.
%
%
Since the seminal works of Reynolds \cite{reynolds1987flocks}, who developed a rule-based motivation for swarming behaviors, and Vicsek \cite{vicsek1995novel} that introduced a simple self-propelled particle model exhibiting a global phase transition, there has been considerable research into the the development of swarming models and analyzing their behavior.  These include many generalizations of the Vicsek model that exhibit different patterns and phase transitions \cite{huepe2004intermittency,gregoire2004onset,aldana2007phase,chate2008modeling,costanzo2018spontaneous} pursued from primarily a physics-based perspective, as well as those that are more biologically motivated \cite{couzin2002collective,romanczuk2009collective,yang2010swarm}.  Swarming has also been widely studied in the engineering community, motivating biomimetic applications in robotics, optimization, and control \cite{kennedy1995particle,passino2005biomimicry,brambilla2013swarm}.

A related phenomenon that has been studied in both the physics and broader communities is the synchronization of coupled oscillators, including the celebrated Kuramoto model \cite{kuramoto1975international} which has spawned much related research (see \cite{strogatz2000kuramoto,acebron2005kuramoto,rodrigues2016kuramoto} for reviews).  The Kuramoto model and self-propelled particle swarms such as \cite{vicsek1995novel} are related \cite{chepizhko2010relation}, and one can view the self-propelled particle model (in two dimensions) as a Kuramoto-like model with time varying network connectivity.  This has led to the use of Kuramoto models in the analysis of swarming behaviors \cite{paley2007oscillator}.
Furthermore, the inter-coupling of oscillator models and swarm dynamics has also been proposed \cite{o2017oscillators,monaco2020cognitive} resulting in additional emergent behaviors due to the interplay between spatial aggregation and phase interactions.

Swarm models are often highly nonlinear, and exhibit nonequilibrium or chaotic behaviors. This can make it challenging to produce closed-form analytical results about stability regimes or other swarm properties. This generally leads to \textit{empirical} analysis of swarming models, where simulations are performed using a range of model parameters to look for ``interesting'' states and phase transitions between them. 
%
The dominant mechanism for empirical analysis of the collective motion in swarming systems is through the use of summary functions and order parameters that attempt to capture global features of the swarm. These can then be analyzed as swarm parameters are swept for the purposes of bifurcation analysis and behavior classification. In the case of \textit{experimental} data, collected through, e.g., video or still imagery, this approach can be used to statistically test a hypothesis,  \cite{schultz2008mechanism,ballerini2008empirical} or to construct or fit a model as in \cite{herbert2011inferring}. The classic example of such an order parameter is the notion of coherence in a swarm, $\frac{1}{N}|\sum_j{e^{i\theta_j}}|$, where $\theta_j$ can refer to either the velocity heading of the $j$th agent as in the Vicsek-style models or a synchronization parameter in the Kuramoto or swarmalator models \cite{vicsek1995novel,couzin2002collective,gregoire2004onset,o2017oscillators}.

Recently, \ac{TDA} \cite{zomorodian2005computing,ghrist2008barcodes,edelsbrunner2008persistent,wasserman2018topological} has been applied to analyze the \textit{topological} structure of collective motion in swarms, i.e., the number of connected ``sub-swarm'' components and the presence holes or voids in the two or three dimensional swarm structure \cite{topaz2015topological,corcoran2017modelling,sinhuber2017phase}.  At the most basic level, these techniques work by defining a series of graphs between agents in the swarm by setting the connectivity by thresholding on inter-agent distances and analyzing how ``persistent'' topological features are when sweeping through this threshold \cite{topaz2015topological,sinhuber2017phase}. A similar concept using spatial density estimates of the swarm was proposed in \cite{corcoran2017modelling}.

In this paper, we show how the emerging field of \ac{GSP} can be used to naturally extend both the analysis of order parameters and \ac{TDA} techniques by incorporating the topological structure of a swarm as a graph, and using graph Fourier analysis to analyze the resulting collective states.
\ac{GSP} builds on its roots in the theory of algebraic signal processing \cite{puschel2008algebraic} and spectral graph theory \cite{chung1997spectral} to analyze functions and signals defined on irregular  domains modeled by graphs and extend techniques from classical signal processing to these domains \cite{shuman2013emerging,sandryhaila2013discrete}. Specifically, we 
%
%
first review some preliminary material from graph theory and \ac{GSP}.
   Next, we discuss different approaches to defining the connectivity graph and defining a graph signal, and relate this to existing approaches to analyzing collective motion in swarms.
   After this, we consider a number of notional (i.e., not generated by simulation) swarm states to provide insight into this approach. 
   We then move on to apply these techniques to a series of swarm simulations from the literature to illustrate their utility, and discuss some finer points of their application.
   Finally, we conclude with a summary discussion, future directions, and potential practical applications.
%

\section{Preliminaries}
%

%
A graph $\mathcal{G}=(\mathcal{V}, \mathcal{E})$ is a collection of vertices $\mathcal{V}=\{\nu_1, \dots, \nu_N\}$ and edges $\mathcal{E}=\{\epsilon_{ij}\}$ between them. A compact description of a graph $\mathcal{G}$ is its adjacency matrix $\mathbf{A}$ where the entries $A_{ij}$ encode the connectivity of $\mathcal{G}$, i.e., $A_{ij}=1$ if $\epsilon_{ij}\in\mathcal{E}$ for unweighted graphs and $A_{ij}=0$.  More generally, a \textit{weighted} graph has $A_{ij}\neq0$ if two vertices are connected and $A_{ij}=0$ otherwise. In this work, we will restrict ourselves to nonnegative weighted, undirected graphs, so that $A_{ij}=A_{ji}\geq 0$ $\forall$ $i,j$ , and assume no self-edges, so that $A_{ii}=0$. If there is a series of edges  connecting all vertices in $\mathcal{G}$, $\mathcal{G}$ is said to be \textit{connected}, and if the graph is disconnected, we will refer to each maximally connected subgraph as a connected component of $\mathcal{G}$.

Related to $\mathbf{A}$ is the combinatorial graph Laplacian $\mathbf{L}=\mathbf{D}-\mathbf{A}$, where $\mathbf{D}$ is a diagonal matrix with $D_{ii}=\sum_jA_{ij}$.  Under the above assumptions on $\mathbf{A}$, $\mathbf{L}$ admits an eigendecomposition $\mathbf{L}=\mathbf{U}\mathbf{\Lambda}\mathbf{U}^\dagger$, where $\mathbf{U}$ is a unitary matrix whose columns are eigenvectors, and $\mathbf{\Lambda}$ is a diagonal matrix of eigenvalues.  An important connection to \ac{TDA} is the fact that the number of zero eigenvalues of $\mathbf{L}$ is equal to the number of connected components in the graph \cite{chung1997spectral}. When the graph is not connected, we will adopt the convention that the vertex ordering is permuted so that the overall graph Laplacian $\mathbf{L}$ is a block diagonal matrix with matrices $\mathbf{L}_\ell$ along the diagonal, where $\mathbf{L}_\ell$ is the Laplacian of the $N_\ell\times N_\ell$ sub-graph for each connected component of $\mathcal{G}$.  With this convention, each eigenvector corresponding to a 0-eigenvalue will be a constant $1/\sqrt{N_\ell}$ on the vertices of that component, and 0 on the remaining vertices \cite{chung1997spectral}.

The normalized graph Laplacian, $\bar{\mathbf{L}}=\mathbf{D}^{-\frac{1}{2}}\mathbf{L}\mathbf{D}^{-\frac{1}{2}}$, is a variation of the combinatorial Laplacian that extends  many of the useful theoretical properties of the combinatorial Laplacian that generally only apply to regular graphs \cite{chung1997spectral}. In particular, it guarantees that the eigenvalues lie in the interval $[0,2]$. These nice theoretical properties have led to its adoption when dealing with irregular graphs, such as the solution to the clustering problem proposed in \cite{shi2000normalized}. However, unlike the combinatorial Laplacian, the eigenvector corresponding to the zero eigenvalue will be non-constant (in fact, the entries will be proportional to $D_{ii}^{1/2}$).

\ac{GSP} studies functions defined on the vertices of graphs, often referred to as graph signals.
Formally, a graph function is a function $\mathbf{f}:\mathcal{V}\to\mathbb{V}^N$ that maps vertices to elements in some vector space, typically $\mathbb{R}^N$ in the \ac{GSP} literature.  In this work, we will consider graph functions defined in the complex numbers ($\mathbb{V}=\mathbb{C}$) for phase and swarm states in two dimensions, as well as $\mathbb{V}=\mathbb{R}^3$ for swarm states in three dimensions.  Finally, we adopt the shorthand notation where $f_i=\mathbf{f}(\nu_i)$. Using this convention, we have that $||f||_2^2=\sum_i ||f_i||^2_2$.


In \ac{GSP} there are many different ways to define fundamentally different \ac{GFT}s using the irregular structure of a graph. These include approaches that use decompositions of the adjacency matrix $\mathbf{A}$, those that use eigendecomposition of the Laplacian (or one of its variations), and those that use variational approaches.  Here, since we assume a positive weighted, undirected graph, we can use properties of $\mathbf{L}$ or $\bar{\mathbf{L}}$ to define a \ac{GFT} that has many desirable properties.  Recall that under our graph assumptions (symmetric and positive weighted), both $\mathbf{L}$ and $\bar{\mathbf{L}}$ are symmetric, positive semidefinite matrices, and as such admit an eigendecomposition $\mathbf{U\Lambda U}^\dagger$ (different for the two Laplacians for a given graph $\mathcal{G}$).  Regardless of the choice of Laplacian, we will define the \ac{GFT} $\hat{\mathbf{f}}\triangleq \mathbf{U}^\dagger \mathbf{f}$, so that the Fourier harmonics are the eigenvectors of the specific Laplacian, and we adopt the convention that the corresponding ``frequency'' of each harmonic is the corresponding eigenvalue $\lambda_j$. For connected graphs, we sort the columns in ascending order of $\lambda_j$. For disconnected graphs, we sort each connected component first in decreasing order of size, followed by increasing order of eigenvalue within the connected components sub-graph Laplacian.

Using this harmonic and frequency convention, many of the usual notions from standard Fourier analysis directly apply to the graph Fourier domain, such as bandlimited signals (i.e., signal concentration in a particular range of graph frequencies) and signal sparsity (i.e., signal concentration in a few graph frequencies).  The structure in a graph signal can be analyzed through the notion of graph filtering a signal $\mathbf{f}$, interpreted in the graph Fourier domain as $\mathbf{U}\mathbf{H}\mathbf{U}^\dagger \mathbf{f}$ for a diagonal matrix $\mathbf{H}$, where the entries $H_{ii}$ are the ``frequency'' response of the filter at the graph frequency $\lambda_i$.  Another \ac{GFT} mechanism for interpreting \ac{GFT} structure of a signal is through the total variation $||\mathbf{f}^\dagger \mathbf{L} \mathbf{f}||_2^2=||\hat{\mathbf{f}}^\dagger\Lambda \hat{\mathbf{f}}||_2^2=\sum_i \lambda_i ||\hat{\mathbf{f}}||_2^2$ (and similarly for $\bar{\mathbf{L}}$), in other words the average graph signal power (with respect to the chosen \ac{GFT}). Normalizing this quantity by the graph signal power is used as a metric of smoothness of a given graph signal. In the following, we will show that many notional and simulated swarm states have graph spectral structure that can be readily determined and exploited by graph filtering and other \ac{GSP} techniques.


\section{GSP Swarm Analysis}
\begin{figure}[!ht]
 
\begin{tikzpicture}






\Vertex[x=5.650,y=3.000]{v0}
\Vertex[x=4.874,y=4.874]{v1}
\Vertex[x=3.000,y=5.650]{v2}
\Vertex[x=1.126,y=4.874,label=$j$]{v3}
\Vertex[x=0.350,y=3.000]{v4}
\Vertex[x=1.126,y=1.126]{v5}
\Vertex[x=3.000,y=0.350]{v6}
\Vertex[x=4.874,y=1.126]{v7}
\Edge[](v0)(v1)
\Edge[](v0)(v7)
\Edge[](v1)(v2)
\Edge[](v2)(v3)
\Edge[](v3)(v4)
\Edge[](v4)(v5)
\Edge[](v5)(v6)
\Edge[](v6)(v7)

\Vertex[x=3,y=3,size=0,color=black]{origin}
\node[below= .1 of origin] (xbar) {$\bar{x}$};
\Edge[style=->](origin)(v3)

\Vertex[x=-1,y=4.2,style={color=white},size=0]{vj}

\Edge[style=->](v3)(vj)

\node[left=of v3,color=black,align=left] (vjlabel) {$\mathbf{v}_j=||\mathbf{v}_j||_2\mathbf{u}_j$\\\phantom{$\mathbf{v}_j$~}$=||\mathbf{v}_j||_2e^{i\phi_j}$};

\node[above = 1 of origin] (rr) {$\mathbf{r}_j=||\mathbf{r}_j||_2e^{i\psi_j}$};

\end{tikzpicture}

\caption{A notional ring state of a swarm for $N=8$, illustrating the topology of a ring graph and various vector and angular states used in the analysis.}

\label{fig:notional_ring}

\end{figure}
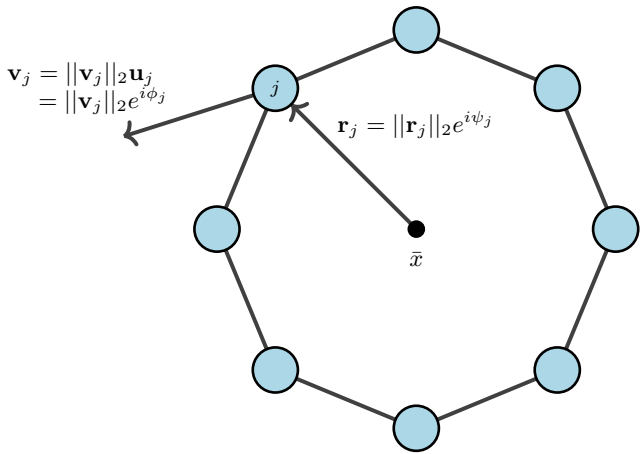

Next, let us define some common notation related to swarms in the abstract sense, without reference to particular dynamical models.  Let $\mathbf{x}_j$, $\mathbf{v}_j$, and $\mathbf{a}_j$ denote the position, velocity, and acceleration vectors, respectively (in $\mathbb{R}^2$ or $\mathbb{R}^3$, as appropriate) of the $j$th agent in a collection of $N$ swarming agents (time index suppressed). Let $\bar{\mathbf{x}}=\sum_j \mathbf{x}_j$ be the center of mass, we define $\mathbf{r}_j=\mathbf{x}_j-\bar{\mathbf{x}}$, and $\mathbf{u}_j$ as the normalized unit vector in the direction of $\mathbf{v}_j$. When considering swarms in two dimensions, we will define the angles $\psi_j$ and $\phi_j$ derived from the relationships $\mathbf{r}_j/||\mathbf{r}_j||_2=(\cos(\psi_j),\sin(\psi_j))^\top$ and $\mathbf{u}_j=(\cos(\phi_j),\sin(\phi_j))^\top$.  The angular variable $\theta_j$ we will use to refer to auxiliary phase variables in swarmalator-based models. Fig.~\ref{fig:notional_ring} shows a notional swarming state in a ring formation with these state variables annotated.

In order to define a \ac{GFT}, one must first define the connectivity of the graph $\mathcal{G}$.  There are several natural ways to do this.  First, when the swarm model interacts using some finite range $R$ or, alternatively, when \ac{TDA} indicates interesting topological structure for some Rips radius $R$ as in \cite{topaz2015topological,sinhuber2017phase}.  In this case, it makes sense to consider $A_{ij}=1$ if $||\mathbf{x}_i-\mathbf{x}_j||<R$, and 0 otherwise.  Here, we will generally use the Euclidean distance, but other options can be used (and should account for any periodic spatial boundary conditions common in the literature). An  alternative to fundamentally discrete approach is to construct a weighted adjacency matrix using some decreasing kernel function of distance, for example $A_{ij}=\exp(-||\mathbf{x}_i-\mathbf{x}_j||^2_2/\sigma^2)$ or $\frac{1}{1+||\mathbf{x}-\mathbf{x}_j||^2_2}$ for $i\neq j$, $A_{ii}=0$. It is also common in the \ac{GSP} literature for these weights to be thresholded below some small value and set to 0. 
A fundamentally different approach was used in \cite{calovi2014swarming} that defined adjacency using Voronoi cells.   
Another method that has been proposed in biological swarms \cite{ballerini2008interaction} is a topological mechanism where e.g., the closest $M$ neighbors are used to define connectivity, but this may not necessarily produce a symmetric graph. 
In this work, we will consider only the first two mechanisms due to their prevalence in the literature, but we stress that the techniques presented here in principle apply to any mechanism used to define a graph, and there exist approaches to performing \ac{GSP} on more general graphs with negative weights and directed edges. 
Finally, we point out that the graph connectivity need not be purely a function of position, and in particular we will look at angular distance $\arccos(\cos(\theta_j-\theta_j))$ as a weighting mechanism for swarmalator models.

As noted above, perhaps the most commonly used approach to analyze collective motion in swarms is to study the velocity alignment using $||\bar{\mathbf{u}}||_2=||\frac{1}{N}\sum_{j} \mathbf{u}_j||_2$, or equivalently $|\frac{1}{N}\sum_j \exp(i\phi_j)|$ in two dimensions. 
Next, consider a graph function $\mathbf{f}_j=\mathbf{u}_j$ where the graph is connected (via e.g., a suitably large $R$ or unthresholded kernel approach).  Then, using the \ac{GFT} defined by the eigendecomposition of the resulting Laplacian we have $\hat{\mathbf{f}}=\mathbf{U}^\dagger \mathbf{f}$, and in particular $\hat{\mathbf{f}}_1=\frac{1}{\sqrt{N}}\sum_j\mathbf{u}_j$ (assuming the columns of $\mathbf{U}$ are sorted by increasing order of eigenvalue).  Thus, $||\hat{\mathbf{f}}_1||_2$ is proportional to the natural alignment order parameter used in the analysis of swarms, i.e., $\bar{\mathbf{u}}$ is analogous to the DC Fourier component of the graph signal.  That said, another natural order parameter to consider in the \ac{GSP} context would be $||\hat{\mathbf{f}_1}||^2_2$, which is analogous to the DC power of the graph signal.  Furthermore, since $\mathbf{U}$ is unitary, $||\hat{\mathbf{f}}||^2_2=||\mathbf{f}||^2_2=\sum_j||\mathbf{u}_j||_2^2=N$ and we can consider the normalized power in the DC harmonic, $||\hat{\mathbf{f}_1}||^2_2/N$, which like $|\bar{\mathbf{u}}|$, is scaled between 0 and 1, representing the extremes of velocity disorder and perfect alignment, respectively.

\begin{figure}[ht!]

\includegraphics[width=\columnwidth]{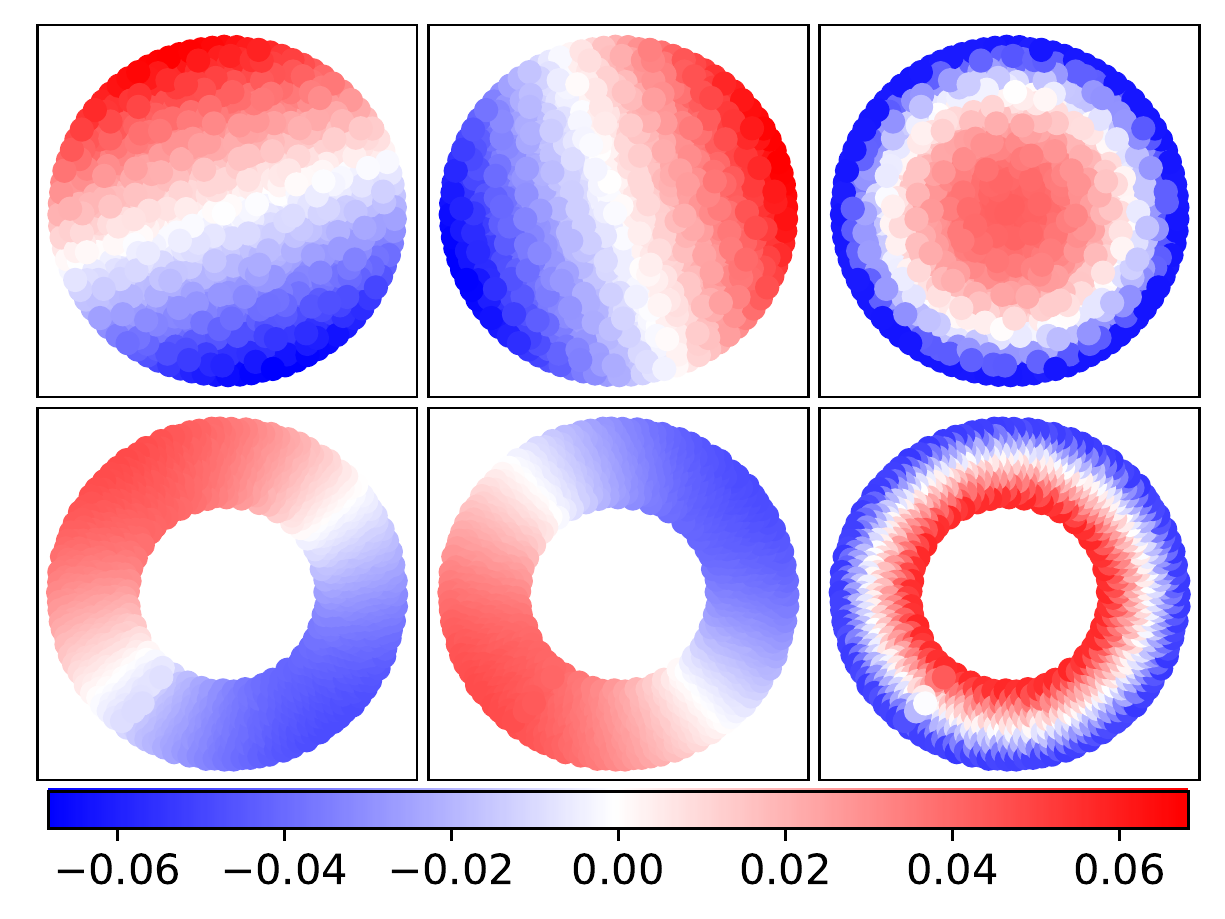}

\caption{Sample Graph Harmonics.  Top row: Graph harmonics for a disk-like state. Bottom row: Graph harmonics for an annular state. Left: Second harmonics showing a standing wave of one period across the graph.  Middle: Third harmonics showing a standing wave of one period across the graph, 90$^\circ$ out of phase with the second harmonic.  Right: Higher order harmonic showing one period of a wave radiating from the center or interior boundary of the swarm.  All positional states were determined from stable states of swarmalator simulations \cite{o2017oscillators}, using $A_{ij}=1/||\mathbf{x_i}-\mathbf{x_j}||_2^2$.  Higher order harmonics include standing waves with higher frequencies and combinations of waved around the swarm structure with waves radiating outward, see Appendix.}
\label{fig:sample_harms}

\end{figure}

Since we consider only unitary \ac{GFT}s which must preserve the overall quantity $||\mathbf{f}||^2_2$ (i.e., the Parseval-Plancherel identity), this raises the question of physical meaning behind signal concentration in these other harmonics. 
For a ring graph, the circulant structure of its Laplacian implies that these higher 
order harmonics are mathematically equivalent to the standard $\cos$ and $\sin$ discrete time Fourier transform harmonics, and is a key motivator in the development of \ac{GSP} from traditional signal processing. Empirically, we find that for the common swarm spatial states that have a roughly disk or annular structure this intuition continues to hold using graph Laplacians as the source for a \ac{GFT}.  In particular, we will focus on the second and third harmonics, which we will show are quite relevant to heading or phase sorted states in swarms.  Figure~\ref{fig:sample_harms} shows sample second and third harmonics for notional disk and annular states, and more examples can be found in \cite{supp}. In addition to harmonics that appear analogous to standard Fourier modes along the closed annular path, there are harmonics that modulate from the inside to the outside of the swarm mass see Figure~\ref{fig:sample_harms} and the Appendix.  
In general, we find that this harmonic behavior is reasonably stable with respect to deformation in the swarm structure and finite size effects, although the particular harmonics and eigenvalues depend on the specific choices that define the underlying adjacency matrix $\mathbf{A}$. In particular, this appears to hold for ``low-frequency'' graph harmonics that capture structural patterns that manifest across the swarm.  

While most of the above discussion focused on the the case where the graph used for the \ac{GFT} is connected, we point out that using the conventions described above that partitions the graph Laplacian $\mathbf{L}$ as a block diagonal of Laplacians $\mathbf{L}_j$ for each connected component of $\mathcal{G}$ results in \ac{GSP} analysis of each connected component independently.  Thus, the first harmonic corresponding to each $\mathbf{L}_j$ can be used to compute velocity alignment for that connected component, and so on for the other concepts discussed above.  In this way, the \ac{GSP}-based analysis presented above naturally and consistently applies to complex swarming behaviors where different ``sub-swarms''  have fundamentally different behaviors (e.g., different aligned or milling groups).

\section{Notional Swarm States}
In this section, we will explore some notional (i.e., not generated by any specific dynamical model) swarm states on ring, annular, toroidal, and other structures.  This analysis provides insight in to the structure apparent in the previous section, and furthermore highlights the flexibility of the GSP approach and demonstrate its relationship to common metrics of swarm structure.

\subsection{Ring States}
%


%
%
%
%

\begin{figure}[!ht]

  \centering

    \includegraphics[width=\columnwidth]{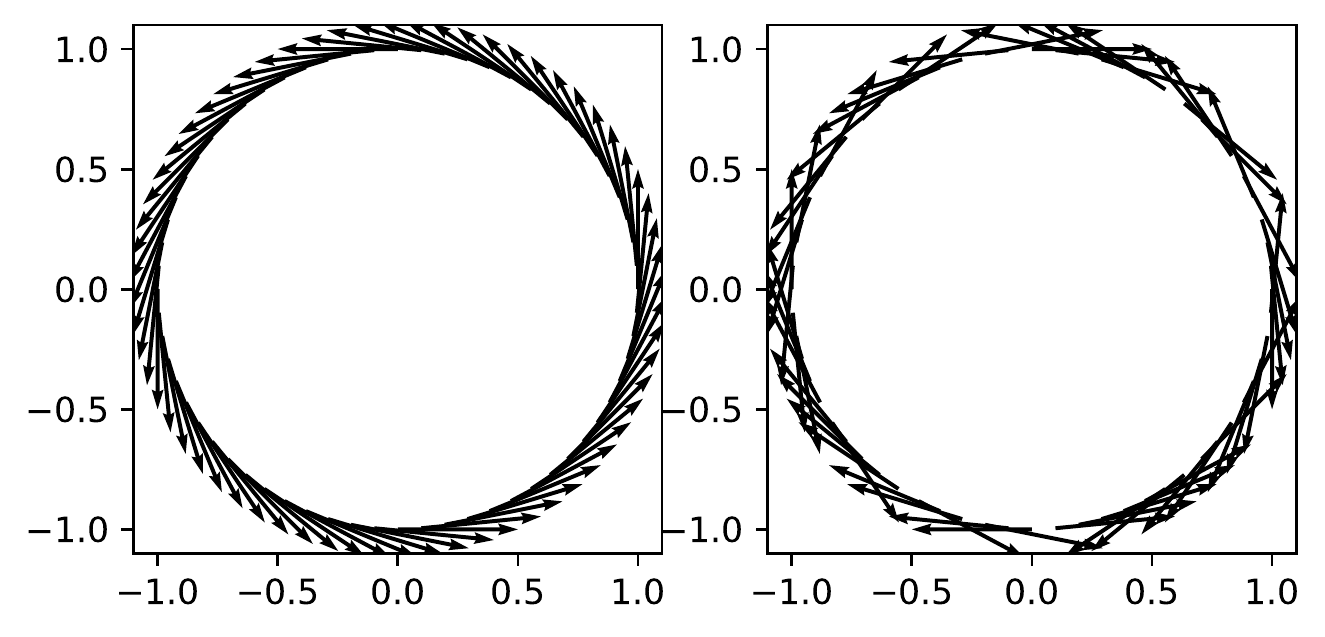}

  \caption{Two notional ring states for swarm size $N$=64. Left: Velocity vectors are aligned in a milling state.  Right: Velocity vectors correspond to a pair of counter-rotating milling states. The positional states are identical for both notional swarm states.}

  \label{fig:two_ring_state}

\end{figure}

\begin{figure}[!ht]

  \centering
  \includegraphics[width=\columnwidth]{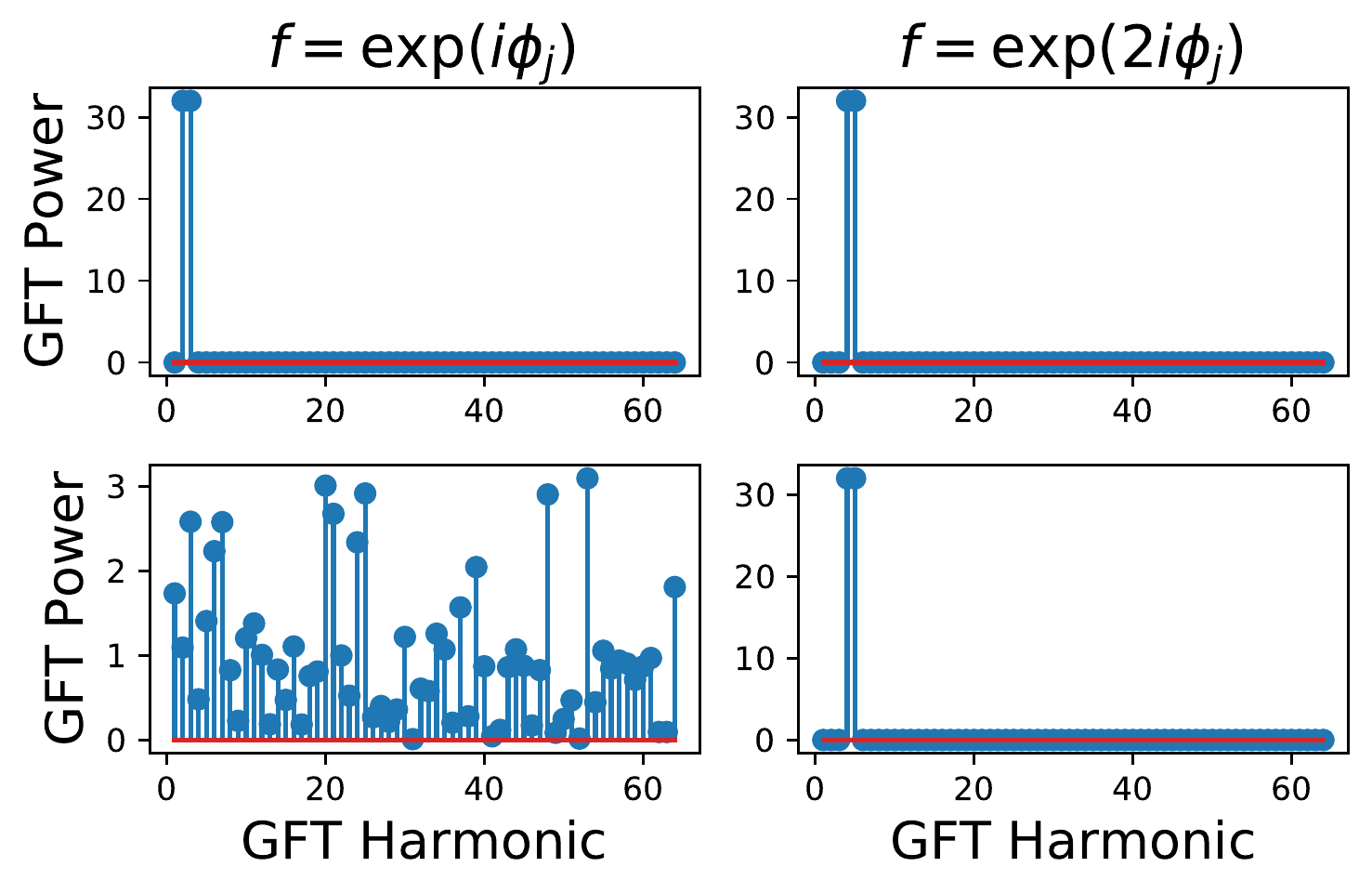}

  \caption{Graph Fourier transforms of the notional swarm states in Fig.~\ref{fig:two_ring_state}, illustrating the difference between milling states that are aligned vs.\ counter-rotating. Top Row: \ac{GFT}s for two graph functions from the aligned milling state. Bottom Row: \ac{GFT}s for two graph functions from the counter-rotating milling state.}

  \label{fig:two_ring_GFTs}

\end{figure}

Fig.~\ref{fig:two_ring_state} shows two different milling states about a ring. The left panel depicts a notional state where all of the velocity vectors align in the same direction along the tangent of the ring. The right depicts a state where the velocity vectors are still tangent to the ring but in both directions along the ring, resulting in a counter-rotating milling state.
As noted above, the Laplacian-based \ac{GFT} for an unweighted ring graph is mathematically equivalent to the standard \ac{DTFT} for real-valued graph signals. 
Thus, in the graph Fourier domain, the aligned ring state has all of its \ac{GFT} power in the second and third harmonics (see Fig.~\ref{fig:two_ring_GFTs}, top left).  As the average velocity of the milling state is exactly zero, we see that there is no \ac{GFT} power in the first (i.e., DC) graph harmonic.  However, the heading angles $\phi_j$ are perfectly tangent and aligned in the same direction along the ring, angular momentum
\begin{equation}
  m_a = \left|\left|\frac{1}{N}\sum_{j=1}^{N}\frac{\mathbf{r}_j\times\mathbf{u}_j}{||\mathbf{r}_j||_2}\right|\right|_2\,,
\end{equation}
that is maximal.
This manifests as concentration in the second and third harmonics, i.e., those corresponding to waves of period one along the ring.  For this ideal case, all of the signal power in these two harmonics, split evenly (at $N/2=32$) between them.

Unlike the aligned milling state, the counter-rotating state does not exhibit any particular structure in the graph Fourier domain (see Fig.~\ref{fig:two_ring_GFTs}, bottom left). In this particular case, we have constructed the notional swarm state so that on average half of the agents are traveling clockwise, and the other half counter-clockwise, resulting in both an average velocity and an angular momentum (and thus concentration in the second and third \ac{GFT} harmonics) that are small. Due to the unitary nature of the \ac{GFT}, we know that the 2-norm must be preserved; the signal power must end up somewhere.  Here, we see this spread appears essentially random, which is consistent with the intuition that the random choice of alignment direction should scramble and spread the spectral content. Despite the apparent lack of spectral structure, it is possible to recover \ac{GFT} structure by considering an alternative graph function.  By using $\exp(i2\phi_j$), i.e., twice the heading angle, as a graph function we see that the transformed signal is concentrated in the fourth and fifth harmonics (Fig.~\ref{fig:two_ring_GFTs}, bottom right).  These harmonics correspond to functions of period 2 along the ring, which is further reinforced by \ac{GFT} analysis of the aligned milling state, which results in a doubling of spectral content (Fig.~\ref{fig:two_ring_GFTs} top right).

\begin{figure}[!ht]

  \centering
  \includegraphics[width=\columnwidth]{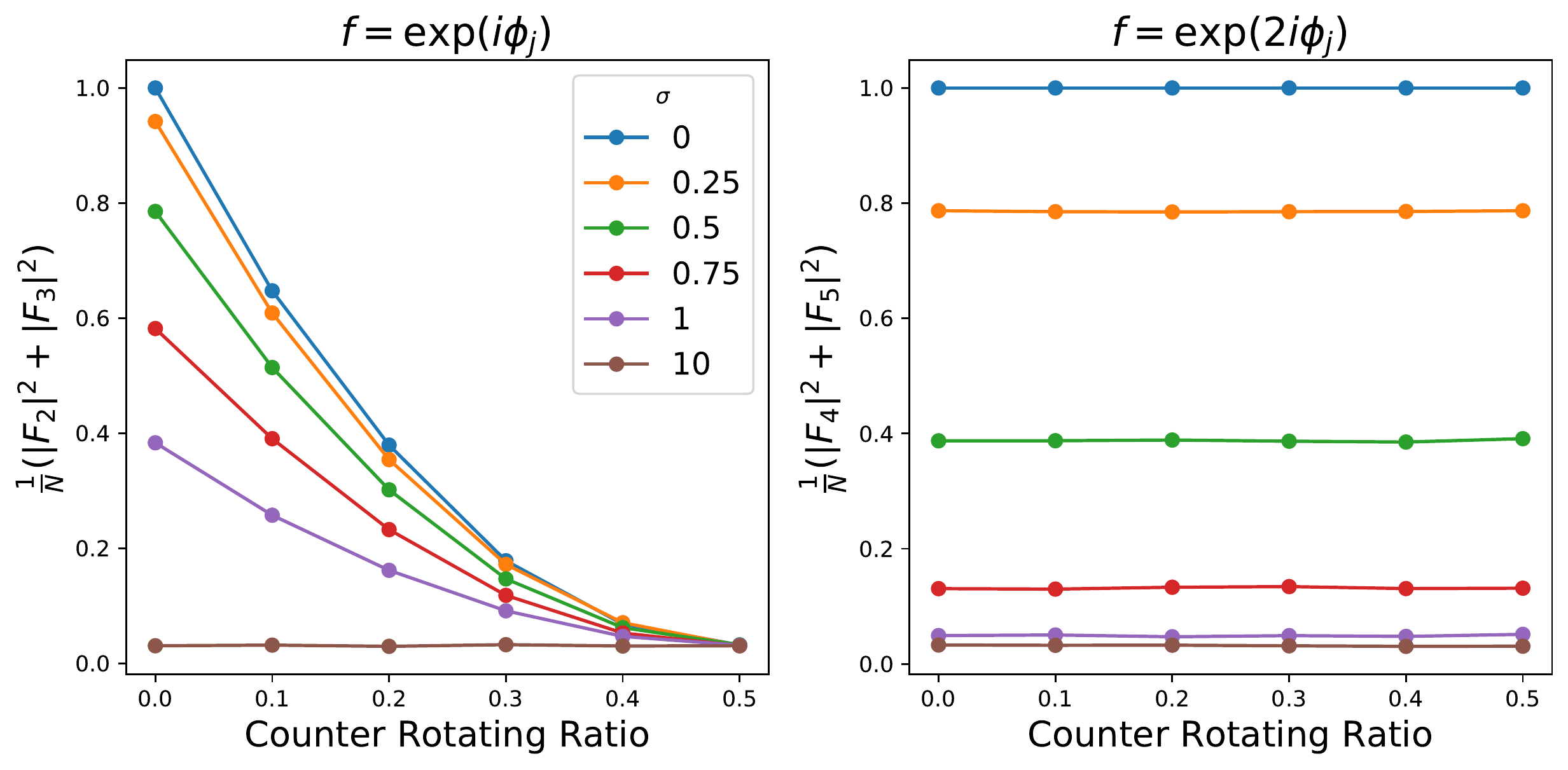}

  \caption{Proportion of power in second and third \ac{GFT} harmonics as a function of the ratio of counter-rotating agents, and the standard deviation $\sigma$ of a wrapped Gaussian angular perturbation to the ideal velocity tangent vector. Left: $\mathbf{f}_j=\exp(i\phi_j)$, showing that the proportion of power these harmonics decreases as either the number of counter-rotating agents increases or the noise variance increases.  Right: $\mathbf{f}_j=\exp(i2\phi_j)$, which is essentially unchanged as the number of counter-rotating agents increases.  This illustrates that these two 
  graph functions recover angular momentum and absolute angular momentum.}

  \label{fig:ring_state_conc_mc}

\end{figure}

The obvious differences in \ac{GFT} power between the graph functions $\exp(i\phi_j)$ and $\exp(i2\phi_j)$ suggests a methodology for analyzing both the overall alignment of the agent motion tangent to the circle and the level of counter-rotation, in much the same way that angular momentum $m_a$ and absolute angular momentum 
\begin{equation}
  M_a = \left|\left|\frac{1}{N}\sum_{j=1}^{N}\frac{||\mathbf{r}_j\times\mathbf{u}_j||_2}{||\mathbf{r}_j||_2}\right|\right|_2\,,
\end{equation}
are traditionally used.  To investigate this we took the same notional swarm position and applied a series of perturbations to the perfectly aligned counter-clockwise rotating milling state (i.e., Fig.~\ref{fig:two_ring_GFTs}, top left). The two perturbations considered were to apply a random Gaussian with mean 0 and standard deviation $\sigma$ to the ideal headings $\phi_j$, and the other to reverse the headings of individual agents with probability $p$, directly controlling the expected ratio of counter-rotation.  The \ac{GFT} power in the second and third harmonics for a set 500 random instances of these perturbations are shown in Fig.~\ref{fig:ring_state_conc_mc}, using both the graph function $\exp(i\theta_j)$ and $\exp(i2\theta_j)$.

These results show that the for the graph function $\exp(i\theta_j)$, the concentration in the second and third harmonics decays monotonically as a function of both deviation from the tangent to the ring (from the Gaussian noise) and consistent rotation direction along the ring. Compare this to the results from the graph function $\exp(i2\theta_j)$ which appears to be strictly a function of the noise in the heading.  Thus, the combination of \ac{GSP} analysis of these two graph functions  it is clear that these two harmonics can be used analogously to order parameters that measure angular momentum and absolute angular momentum (and of course, the first graph harmonic still captures the angular coherence of the swarm).

\subsection{Annular States}
\begin{figure}[!ht]

  \centering
  \includegraphics[width=\columnwidth]{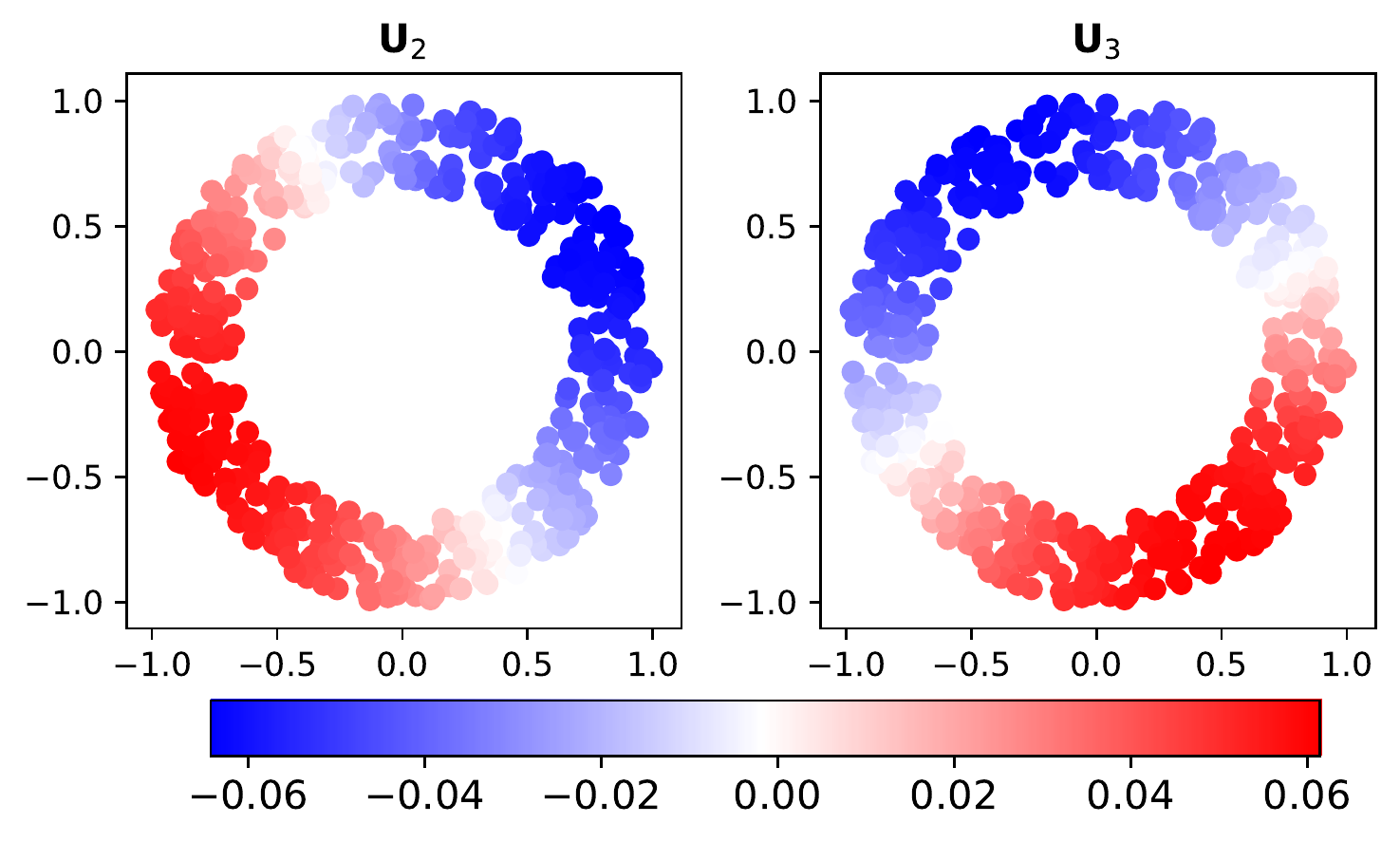}

  \caption{Second and third \ac{GFT} harmonics for a randomly generated notional annular state.$N=512$ agents were placed  uniformly at random on an annulus with outer diameter of one unit and inner diameter of $\frac{2}{3}$.  The adjacency matrix was defined by $A_{ij}=1$ if $||x_i-x_j|_2|<\frac{1}{4}$ and 0 otherwise ($i\neq j$), $A_{ii}=0$. The combinatorial Laplacian $\mathbf{L}$ was used to define the \ac{GFT}. Note that these harmonics have period one with respect to the annular structure and are approximately $90^\circ$ out of phase.}

  \label{fig:annular_harmonics}

\end{figure}

The intuition built in the previous section using a ``perfect'' ring state carries over into less-regular states, due in part to the robustness of the topological underpinnings of this form of \ac{GSP} to perturbations in the physical positions of the swarming agents.  Here, we show that the \ac{GSP} analysis of swarm states that are approximately annular yields similar results to the ring state analyzed above (see also the motivating example).  Again, we proceed with a notional annular state to enforce the idea that this analysis is independent of the dynamical model being studied, excepting prior knowledge about the interaction range. Here, we place $N=512$ agents uniformly at random on an annulus with outer diameter of one unit and inner diameter of $\frac{2}{3}$.  We then define a graph with $A_{ij}=1$ if $||x_i-x_j||<\frac{1}{4}$ and 0 otherwise ($i\neq j$), with $A_{ii}=0$.  Fig.~\ref{fig:annular_harmonics} shows an example of such a positional state and its corresponding combinatorial Laplacian \ac{GFT} harmonics $\mathbf{U}_2$ and $\mathbf{U}_3$.

As is the case with the notional ring state, these harmonics correspond to basis elements that vary with period one across the closed annular path, and are roughly $90^\circ$ out of phase, thus generalizing the standard $\cos(\cdot)$ and $\sin(\cdot)$ harmonics. Unlike the perfect ring state, where higher order \ac{GFT} harmonics correspond to higher frequency periodic structure along the ring, higher order harmonics of an annulus can correspond to higher frequency oscillations along the closed annular path or variations on the inner/outer axis of the annulus, as well as combinations of both (see Fig.~\ref{fig:sample_harms} and \cite{supp}).  Using $\psi_j$, the angle of each notional position, we can define an ideal clockwise motion along the annulus as $\psi_j+\frac{\pi}{2}$, and similarly a counter-clockwise direction as $\psi_j-\frac{\pi}{2}$.  Using either of these as a graph signal results in nearly all of the signal concentration in the first and second \ac{GFT} harmonics, but unlike the ring state the power in the two harmonics are generally not equal, and some of the other harmonics will have small amounts of residual \ac{GFT} power.  For example, due to the random nature of the swarm positions, the swarm center of mass $\bar{x}$ will have non-zero norm and thus it is highly unlikely that the individual headings will cancel out perfectly.

\begin{figure}[!ht]

  \centering
  \includegraphics[width=\columnwidth]{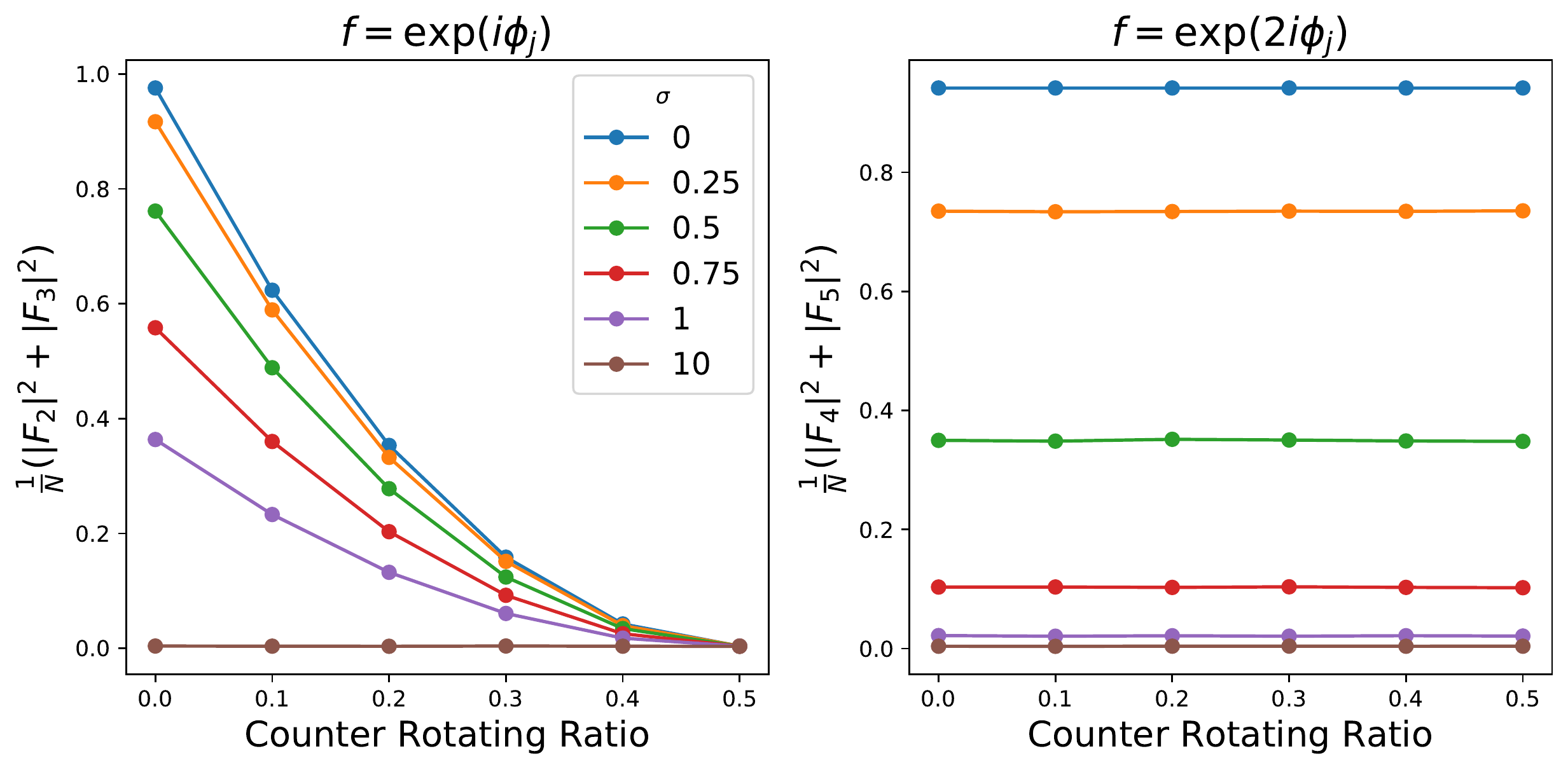}

  \caption{Proportion of power in second and third \ac{GFT} harmonics (see Fig.~\ref{fig:annular_harmonics}) as a function of the ratio of counter-rotating agents for an annular states, and the standard deviation $\sigma$ of a wrapped Gaussian angular perturbation to the ideal velocity tangent vector. Left: $f=\exp(i\phi_j)$, showing that the proportion of power these harmonics decreases as either the number of counter-rotating agents increases or the noise variance increases.  Right: $f=\exp(i2\phi_j)$, which is essentially unchanged as the number of counter-rotating agents increases.  This illustrates that these two 
  graph functions recover angular momentum and absolute angular momentum, in much the same way as the ring state in Fig.~\ref{fig:ring_state_conc_mc}.}

  \label{fig:annular_state_conc_mc}

\end{figure}

Despite the random positions introducing some non-idealities from the perfect ring state, we find that a similar analysis of deviations from the ideal clockwise (or counter-clockwise) headings defined using $\psi_j$ produces nearly identical results as in the ring state (see Fig.~\ref{fig:annular_state_conc_mc}).  Again, the concentrations in the two harmonics is overall slightly less than in the ring case.  However, the general trends relating the transforms of the two graph functions holds, indicating again that these harmonics are capturing essentially the same information as angular momentum and absolute angular momentum.

\subsection{Torus States}
\begin{figure}[!ht]

  \centering
  \includegraphics[width=\columnwidth]{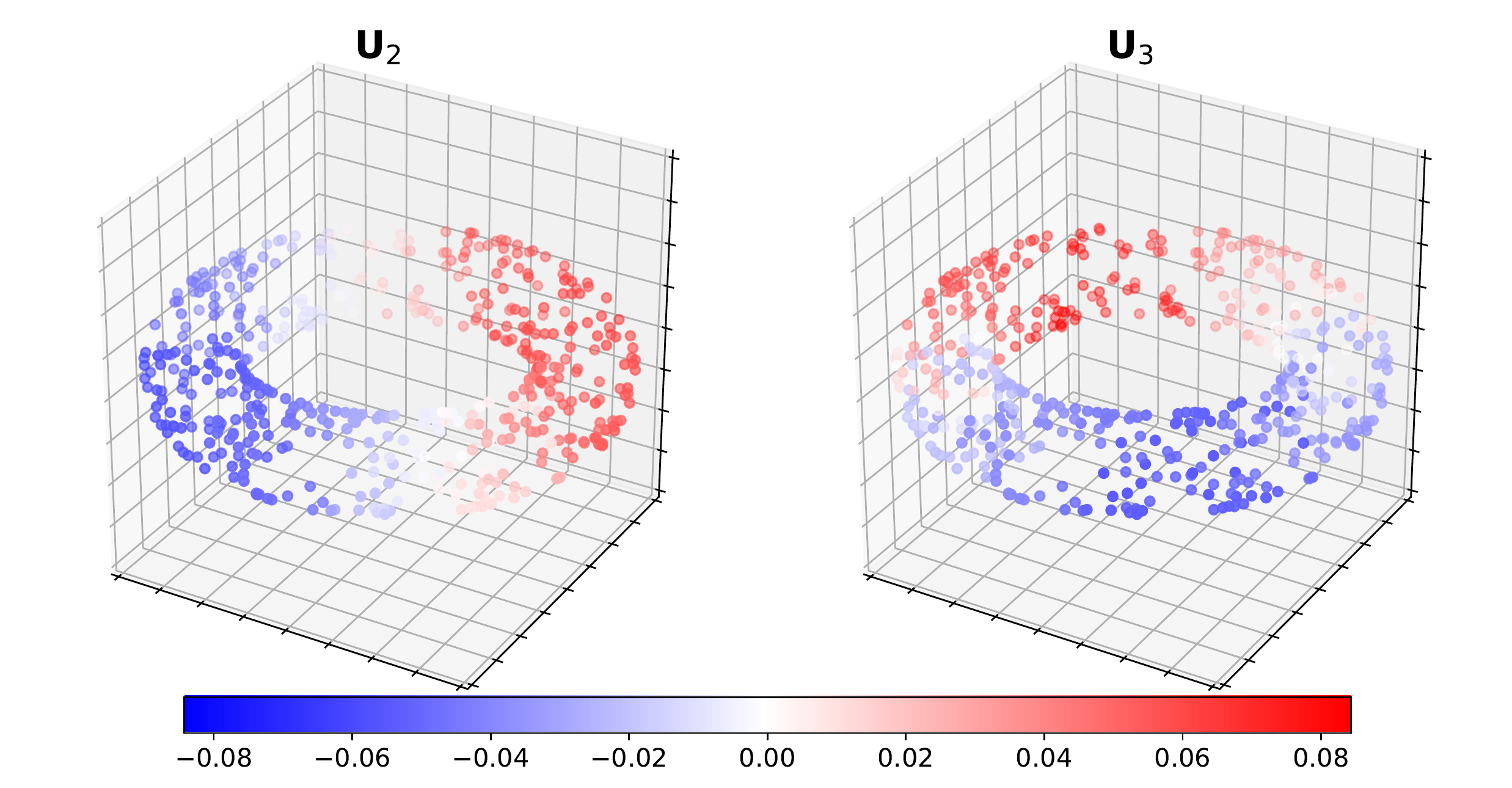}

  \caption{Second and third \ac{GFT} harmonics for a randomly generated notional torus state.$N=512$ agents were placed  uniformly at random on a torus with toroidal radius one unit and poloidal radius $\frac{1}{4}$ units.  The adjacency matrix was defined by $A_{ij}=1$ if $||x_i-x_j||<\frac{1}{4}$ and 0 otherwise ($i\neq j$), $A_{ii}=0$. The combinatorial Laplacian $\mathbf{L}$ was used to define the \ac{GFT}. Note that these harmonics have period one with respect to the toroidal direction and are approximately $90^\circ$ out of phase.}

  \label{fig:torus_harmonics}

\end{figure}

Torus states are three dimensional analogues of the ring and annular states observed in two dimensions, and occur in swarming models \cite{couzin2002collective} and experimental data \cite{parrish2002self} To gain some initial insight into this more complex structure, we constructed a notional torus swarming state with major (``toroidal'') radius  of one distance unit, and minor (``poloidal'') radius of $\frac{1}{4}$. We then sampled this surface uniformly to create a notional swarm position state for a swarm of size $N=512$. As with the notional annular state in the previous section, 
we defined a graph with $A_{ij}=1$ if $||x_i-x_j||<\frac{1}{4}$ and 0 otherwise ($i\neq j$), with $A_{ii}=0$.  Fig.~\ref{fig:torus_harmonics} shows an example of such a positional state and its corresponding combinatorial Laplacian \ac{GFT} harmonics $\mathbf{U}_2$ and $\mathbf{U}_3$. These harmonics bear a striking resemblance to the annular harmonics in Fig.~\ref{fig:annular_harmonics} in that they both have period one (along the toroidal direction) and are approximately $90^\circ$ out of phase.

\begin{figure}[!ht]

  \centering
  \includegraphics[width=\columnwidth]{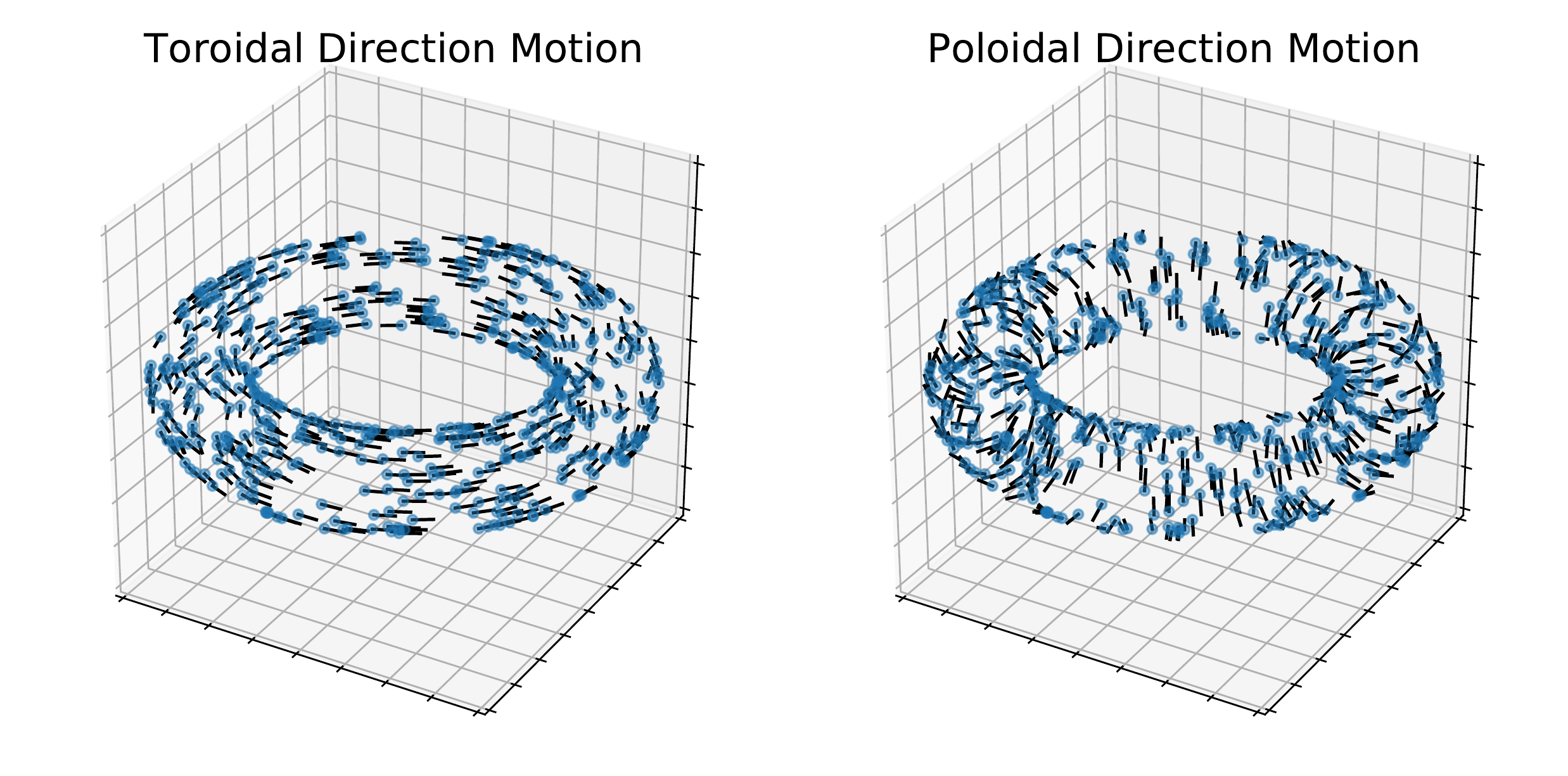}

  \caption{Example motion states along a torus. Left: Coherent motion in the toroidal direction. Right: Coherent motion in the poloidal direction.}

  \label{fig:torus_motion}

\end{figure}

\begin{figure}[!ht]

  \centering
  \includegraphics[width=\columnwidth]{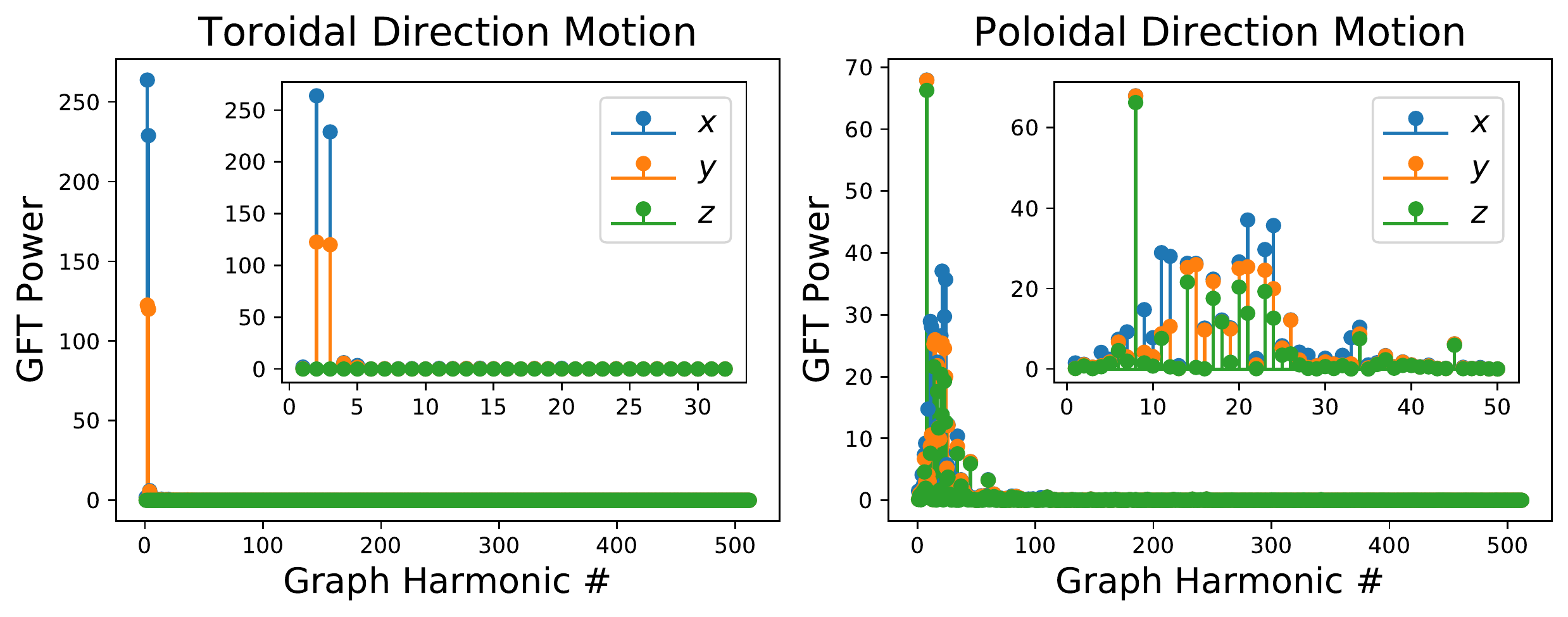}

  \caption{Stem plots of \ac{GFT} power for motion along toroidal direction (left) and poloidal direction (right). \ac{GFT} power is shown decomposed into the $x$, $y$, and $z$ dimensions, with the height of each color indicating the contribution from that dimension in a stacked fashion.  Since the toroidal direction motion has no $z$ components, its contribution to the \ac{GFT} power is zero, and the \ac{GFT} structure is essentially the same as in the annular states.  The poloidal direction is more complex and while still a ``low-frequency'' signal, has higher frequency contributions from all three axes (notably, their are negligible contributions in the first through third harmonics).}

  \label{fig:torus_GFT}

\end{figure}

As was the case with the notional annular structure, we can use the randomly generated position states to define some natural collective motions that align with these graph harmonics.  Rotation along the toroidal direction can be defined in an essentially identical fashion to the annular state once one observes that any velocity in the $z$ direction should be zero in this case.  Thus, coherent motion along the toroidal direction is determined by adding $\pm\frac{\pi}{2}$ to the angle determined by the position projected into the $x-y$ plane.  One such perfectly aligned state is shown in Fig.~\ref{fig:torus_motion}. As one might expect, \ac{GSP} analysis of this (now three-dimensional) graph signal shows nearly all signal power lies in the second and third harmonics (Fig.~\ref{fig:torus_GFT}).  Additionally, we have decomposed the signal power individually into $x$, $y$, and $z$, components, and we see that the $x$ and $y$ dimensions each contain roughly half of the overall signal power, with no power in the $z$ dimension. As noted in the previous section, the randomness-induced irregularity in position results in unequal powers in the two harmonics, and furthermore between the spatial dimensions.

In addition to collective motion along the toroidal direction, another direction of motion to consider is motion along the poloidal axis.  Again, this can be constructed from the notional positional state by looking at the angle of displacement of an agent's position from the toroidal circle in the plane spanned by $\mathbf{r}_j$ and $z$. One such state is shown in Fig.~\ref{fig:torus_motion}, where the collective poloidal motion depicts agents moving ``upward'' (i.e., positive $z$ motion) along the outside of the torus and downward along the inside.
In principle, this motion is topologically equivalent to the coordinated motion along the toroidal direction, but from the more geometric perspective offered by \ac{GSP}, toroidal axis motion should be smoother (i.e., lower frequency) than the poloidal axis motion which varies much faster as a function of distance, not only along the poloidal direction, bus also along the toroidal.  This is reflected in the \ac{GFT} power decomposition shown in Fig.~\ref{fig:torus_GFT} that does not exhibit as clear of a pattern as the toroidal axis motion, but is still decidedly low frequency over all.  In particular, we see that there is little spectral content in the first three harmonic, indicating that the toroidal and poloidal motions are orthogonal, as expected.  Additionally, unlike the toroidal motion, poloidal motion has content in $x$, $y$, and $z$ dimensions, and there is considerable variability in the distribution in each harmonic across these three dimensions. 

\begin{figure}[!ht]

  \centering
  \includegraphics[width=.8\columnwidth]{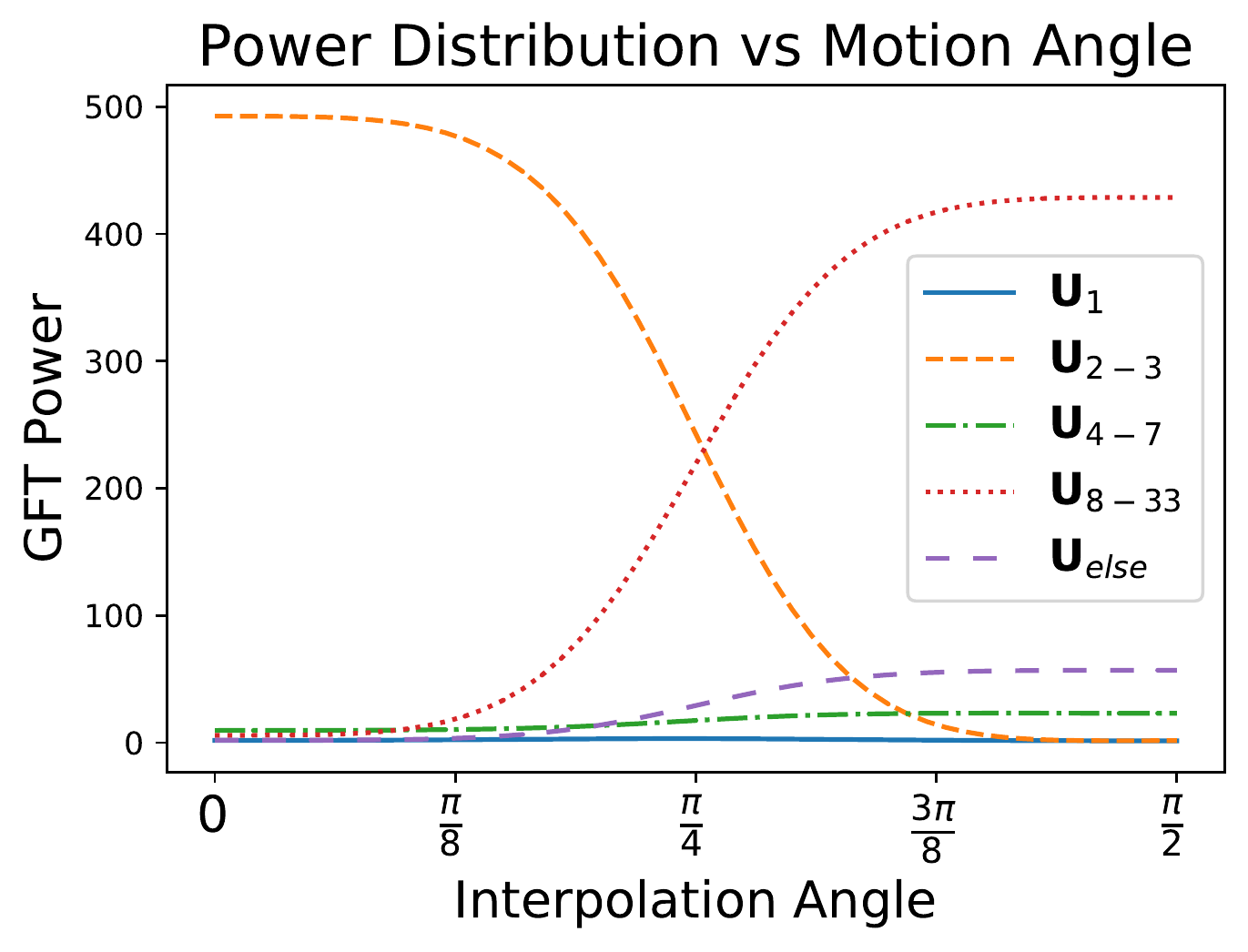}

  \caption{Transition of power in \ac{GFT} harmonics as the motion along the notional torus shape is interpolated from motional along the toroidal direction to motion along the poloidal direction. Since the \ac{GFT} is linear, a linear combination of the two perpendicular directions results in a linear combination in the \ac{GFT} power.}

  \label{fig:torus_interp}

\end{figure}

As a further example of how \ac{GSP} can be used to analyze collective motion, we consider aligned motion along the torus that sits between the toroidal and poloidal directions, tracing out a helical pattern along torus.  The toroidal direction and poloidal direction define orthogonal axes at each notional position on the torus, so we can define a notional helical velocity state of angle $\varphi$ by assigning a heading of $\phi_i$ to each agent that is the linear combination of the notional toroidal and poloidal motions, weighted by $\cos(\varphi)$ and $\sin(\varphi)$, respectively.  From the \ac{GSP} perspective, since the graph signal corresponding to the notional helical motion is a linear combination of the graph signals for toroidal and poloidal motion, and a \ac{GFT} is linear by construction, we should see that the \ac{GFT} of the helical motion is a linear combination of the respective \ac{GFT}s of the torioidal and poloidal states. This is illustrated in Fig.~\ref{fig:torus_interp} which shows a smooth transfer of \ac{GFT} power from toroidal motion ($\varphi=0$) to poloidal motion $(\varphi=\frac{\pi}{2}$).

While the above results focused on ``hollow'' toroidal structures, we find that similar results are produced for notional sold torus states, despite the fundamental differences in their topology \cite{supp}. The graph harmonics and analysis of the toroidal direction of motion are essentially the same. However, the \ac{GFT} signature of the poloidal is less structured and overall less concentrated towards low frequency (and thus the shift between the two as in Fig.~\ref{fig:torus_interp} is not clear).  This is not particularly surprising since poloidal motion in a filled torus should result in even less alignment between neighbors. Consider, for example, what does poloidal direction motion look like at the ``center'' of the torus.


\subsection{$\mathbf{r}$ as a graph signal}
\begin{figure}[ht]

  \centering
  \begin{tabular}{cc}

  \includegraphics[width=.5\columnwidth]{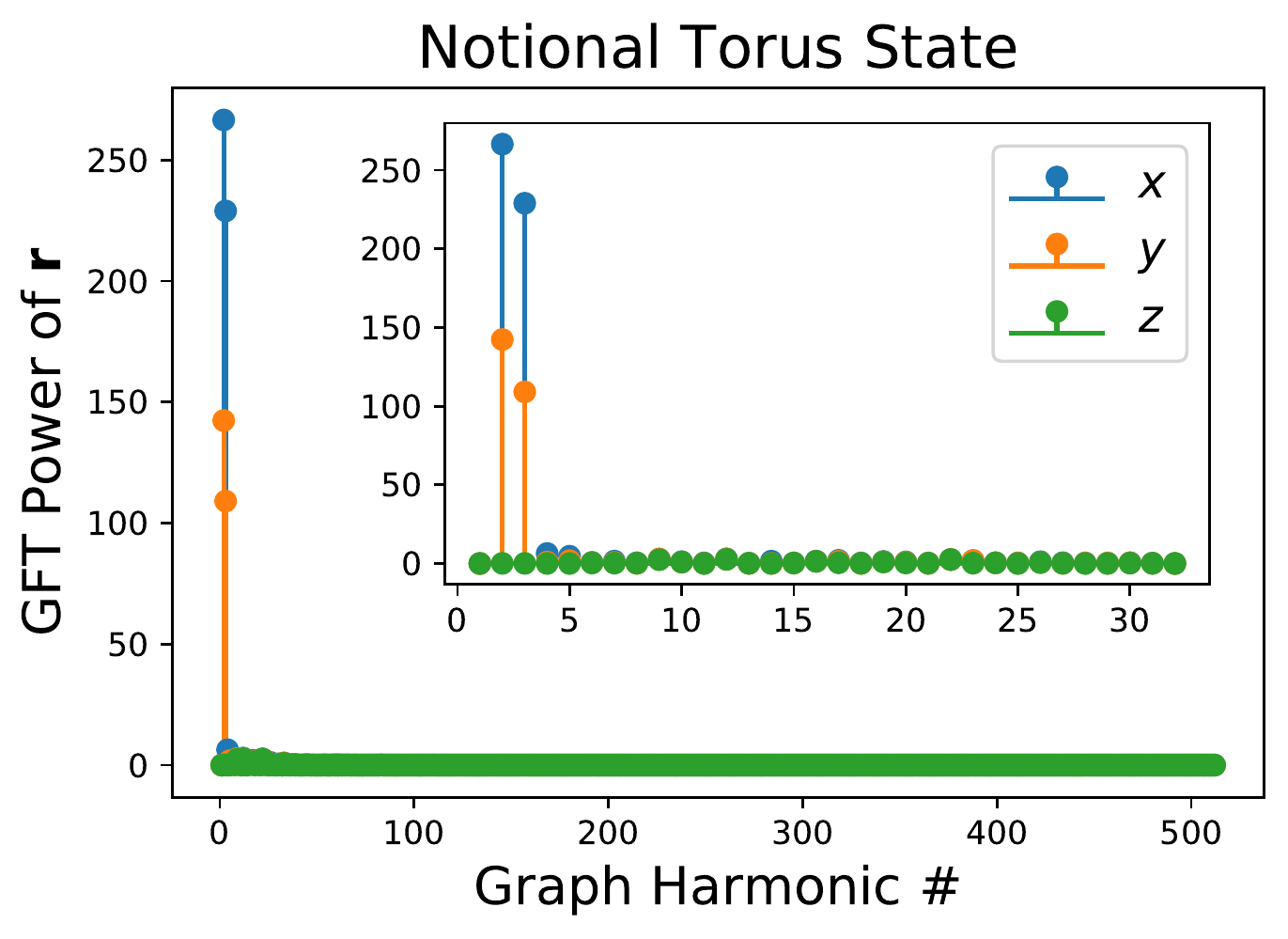} &
  \includegraphics[width=.5\columnwidth]{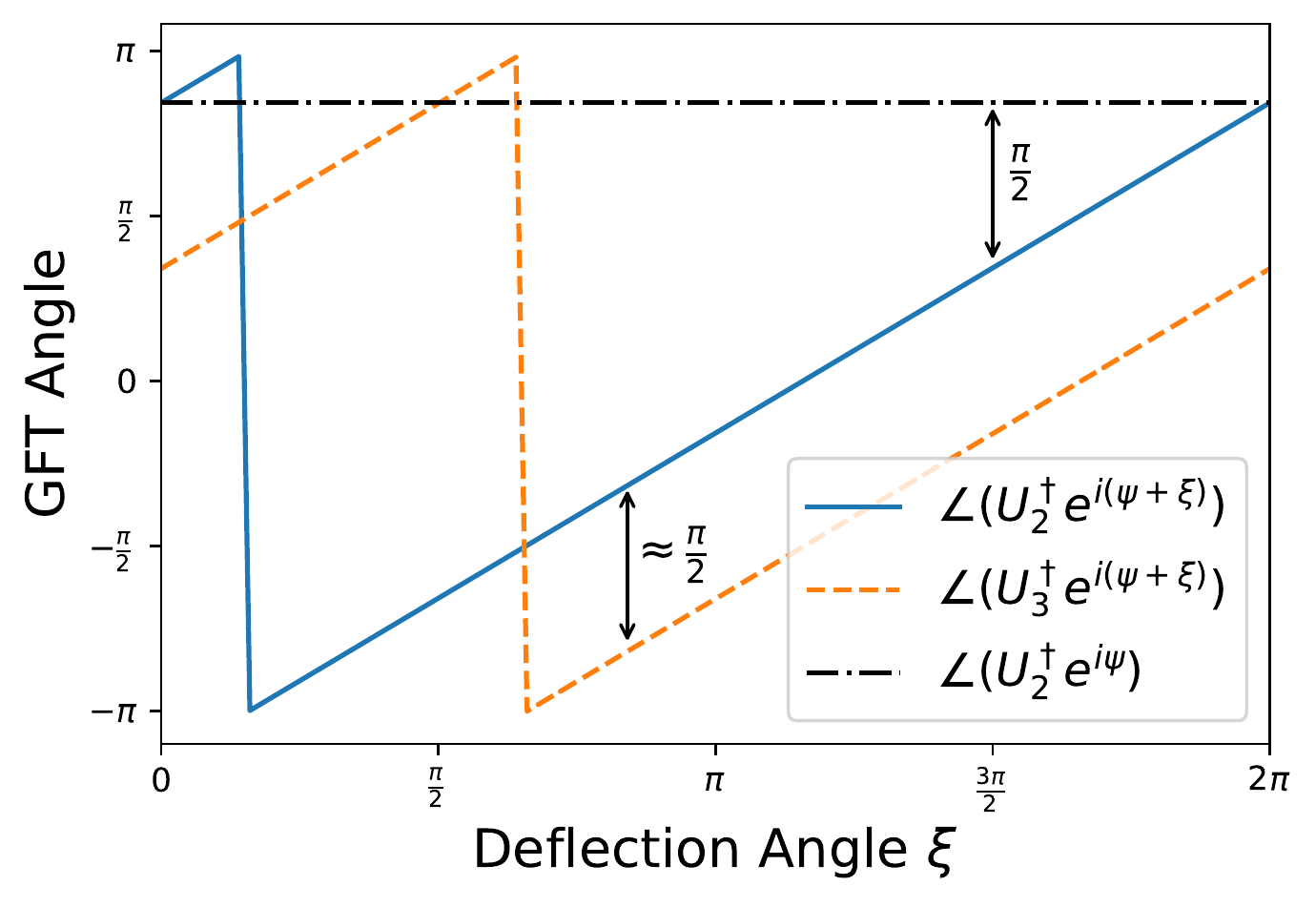}
  \end{tabular}

  \caption{Left: \ac{GFT} power of the the notional torus position state $\mathbf{r}$. Right: Illustration of the linearity of the \ac{GFT} by rotating each position $\mathbf{r}_j$ by an angle $\xi$ and showing how the impacts on the phase of harmonics two and three for the notional annular state. }
  \label{fig:r_sig}

\end{figure}

Thus far, the notional swarming states that we have considered are symmetric in a certain sense with respect to $\mathbf{r}$.  When we treat $\mathbf{r}$ as our graph signal of interest (as opposed to $\mathbf{u}$), we see that the \ac{GFT} concentration is essentially identical to the response of $\mathbf{u}$ for the aligned motion along the ring, aligned motion along the annulus, and toroidal direction motion on the torus (see e.g., Fig.~\ref{fig:r_sig}, left).  This can be understood through the linearity of the \ac{GFT}. Since, by construction, the velocities in these notional states are defined by rotating each $\mathbf{r}_i$ by the same linear rotation map, the impacts in the \ac{GFT} domain will only show up in the \text{phase}, which in these examples amounts to the transfer of signal between the dimensions of $\hat{\mathbf{r}}_i$.

To see this concretely,  Fig.~\ref{fig:r_sig} (right) shows the angle of the \ac{GFT} of the signal $\mathbf{f}_j=\exp(i(\psi_j+\xi))$ at the second and third harmonics, for different angles $\psi$. While the power in each harmonic remains fixed for each $\xi$, we see that the angle of the two harmonics tracks this shift with the two harmonics being approximately $\frac{\pi}{2}$ radians out of phase.  In particular, when $\xi=\pm\frac{\pi}{2}$ (corresponding to perfect tangent motion to the annular structure), we see that $\mathbf{U}_2$ is out of phase by the same amount.  Similarly, we could apply an identical rotation in three dimensions to each component of $\mathbf{r}$ in the notional torus state and see a similar effect. However, interpolation between the toroidal and poloidal directions cannot be represented as identical rotations of each component of $\mathbf{r}$, and this is why signal content leaves harmonics two and three in Fig.~\ref{fig:torus_interp}.



\begin{figure}[ht]

    \centering
    \includegraphics[width=\columnwidth]{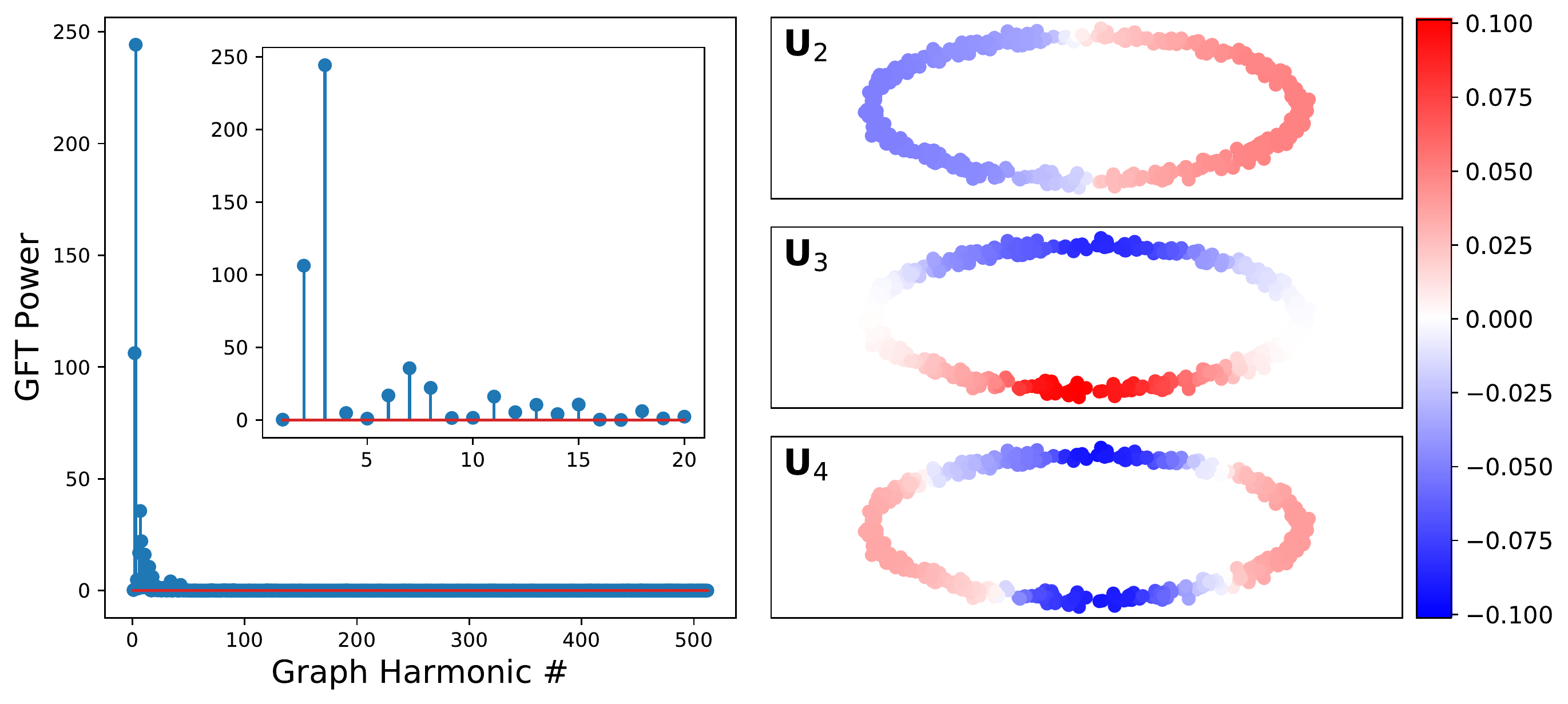}

    \caption{GFT Power of the signal $\mathbf{u}$ and sample harmonics for an an oblong ring-like shape using a disk cut-off and the normalized Laplacian. When applied to the graph signal $\mathbf{r}$, the signal is approximately $86\%$ concentrated in $\mathbf{U}_2$ and another $7.5\%$ in $\mathbf{U}_3$.  Unlike the ring, annular, and toroidal direction examples, whose angular momentum $m_a\approx 1$, tangent motion along this curve produces $m_a\approx 0.75$.}
    \label{fig:oval_GFT_disk}

\end{figure}

Given the above connections between the spectra of $\mathbf{r}$ and $\mathbf{u}$, we next investigate how much the role of perfect symmetry of $\mathbf{r}$ is playing in the above results. To do this, we created additional ring-like state using the oblong closed Lissajous curves defined by $(x(t),y(t))= (3\cos(t), \sin(t))$ for $t\in[0,2\pi)$. For each of $512$ agents, we picked an angle uniformly at random to determine its nominal position along the curve and then perturbed this in each dimension by a zero-mean Gassian with standard deviation $0.05$.  The notional velocities $\mathbf{v}_j$ are set to be normalizations of the derivative of this curve $(-3\sin(t), \cos(t))^\top$.  Using a ``disk-based'' cutoff of $||\mathbf{x}_j-\mathbf{x}_k||_2^2<0.1$ and the combinatorial Laplacian, we find that the angular momentum $m_a\approx 0.75$, far less than the previously considered notional states, despite the fact that there is stong concentration in the \ac{GFT} power for both $\mathbf{u}$ (in harmonics 2 and 3, see Fig.~\ref{fig:oval_GFT_disk}) and $\mathbf{r}$ ($\approx86\%$ in the second harmonic).  Of course, the angular momentum should be less in this example, as the momentum along the longer portion of the distorted ring is more linear than the shorter portion.

As was the case with the notional annulus state, we see that the first two harmonics are $90^\circ$ out of phase, and here they split the oblong ring along the minor and major axes, respectively. However, unlike the ring state, annular state, and toroidal direction state, here we have both a massive imbalance in the power in the two harmonics \textit{and} the imbalances are in the opposite direction for $\mathbf{u}$ and $\mathbf{r}$. This would appear to be a fundamental difference between this state and the more symmetric states considered above.

As the harmonics in Fig.~\ref{fig:oval_GFT_disk} above are not as symmetric as we might hope (see for example the difference in length of the colored regions for $\mathbf{U}_4$ in Fig.~\ref{fig:oval_GFT_disk}), we considered graph definitions using $A_{jk}=\exp(-||\mathbf{x}_j-\mathbf{x}_k||_2^2/\sigma^2)$ where $\sigma^2 = \frac{1}{N(N-1)}\sum_{j,k}||\mathbf{x}_j-\mathbf{x}_k||_2^2$ and \ac{GFT}s using the corresponding combinatorial Laplacian, as well as normalized versions of the Laplacian for the original disk-based graph definition and the weighted one.  As with the case above, we find that the vast majority ($\approx90\%$) of the \ac{GFT} power of $\mathbf{r}$ is concentrated in their respective second harmonics for all three additional cases. For the disk-based graph, using normalized Laplacian we see another $\approx8\%$ in $\mathbf{U}_3$.  Oddly, the weighted graph with combinatorial Laplacian has negligible power in all remaining bands.  For the signal $\mathbf{u}$, the disk-based graph using the normalized Laplacian is very similar to the combinatorial Laplacian albeit with slightly more concentration ($\approx77$  vs.\ $\approx68$) in harmonics two and three.  Again, as with $\mathbf{r}$ the exponentially weighted graph is dominated by a single harmonic, containing a dismal $24\%$ of the overall power.

\begin{figure}[ht]

    \centering
    \includegraphics[width=\columnwidth]{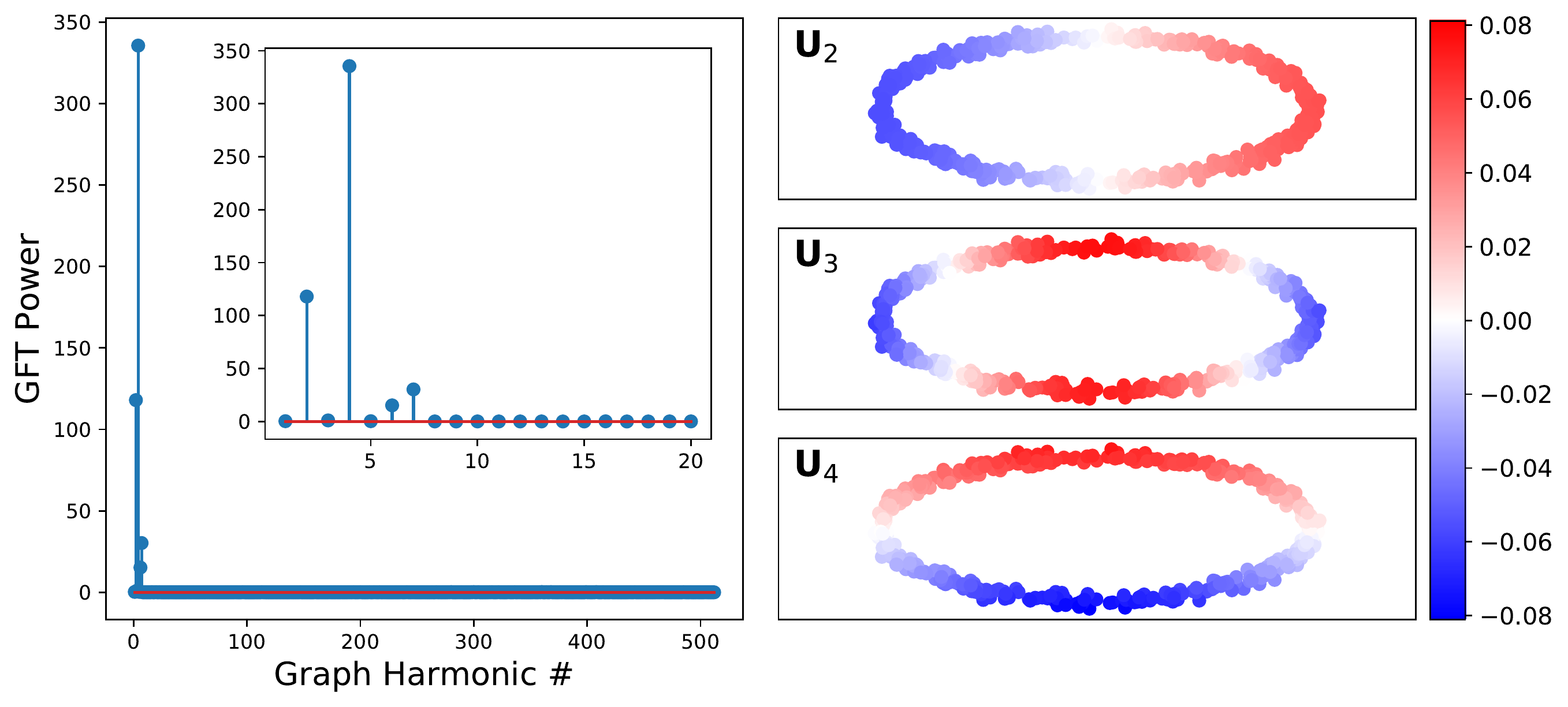}

    \caption{GFT Power of the signal $\mathbf{u}$ and sample harmonics for an oblong ring-like shape using a squared exponential weight and the normalized Laplacian. When applied to the graph signal $\mathbf{r}$, the signal is approximately $90\%$ concentrated in $\mathbf{U}_2$ and another $9\%$ in $\mathbf{U}_4$.}
    \label{fig:oval_GFT_exp}

\end{figure}

The weighted graph with the normalized Laplacian has several spectral features worth noting.  Firstly, its harmonics appear to be more symmetric with respect to the structure of the oblong state, see Fig.~\ref{fig:oval_GFT_exp}, and furthermore, it indicates that the frequency ordering of the harmonics is somewhat perturbed as compared to the other three combinations.  To this point, the \ac{GFT} power of $\mathbf{r}$ is very similar to the the disk-based adjacency with the normalized Laplacian (90\% and 9\%), but with secondary concentration in $\mathbf{U}_4$, as we might expect given the similarities in Fig.~\ref{fig:oval_GFT_disk} and \ref{fig:oval_GFT_exp}.  With respect to $\mathbf{u}$ we see considerably more spectral concentration in the two dominant harmonics (here $\mathbf{U}_2$ and $\mathbf{U}_4$ than the other cases, and these are almost in a 3:1 ratio (i.e., the exact ratio of the major to the minor axis). Overall, we conjecture that the normalized Laplacians are outperforming the combinatorial ones due the variations in density along the curved state, and this is especially relevant for the weighted case, where the agents near the poles of the minor axis are much closer to the rest of the agents, on average, than those near the poles of the major axis.

\begin{figure}[ht]

    \centering
    \includegraphics[width=\columnwidth]{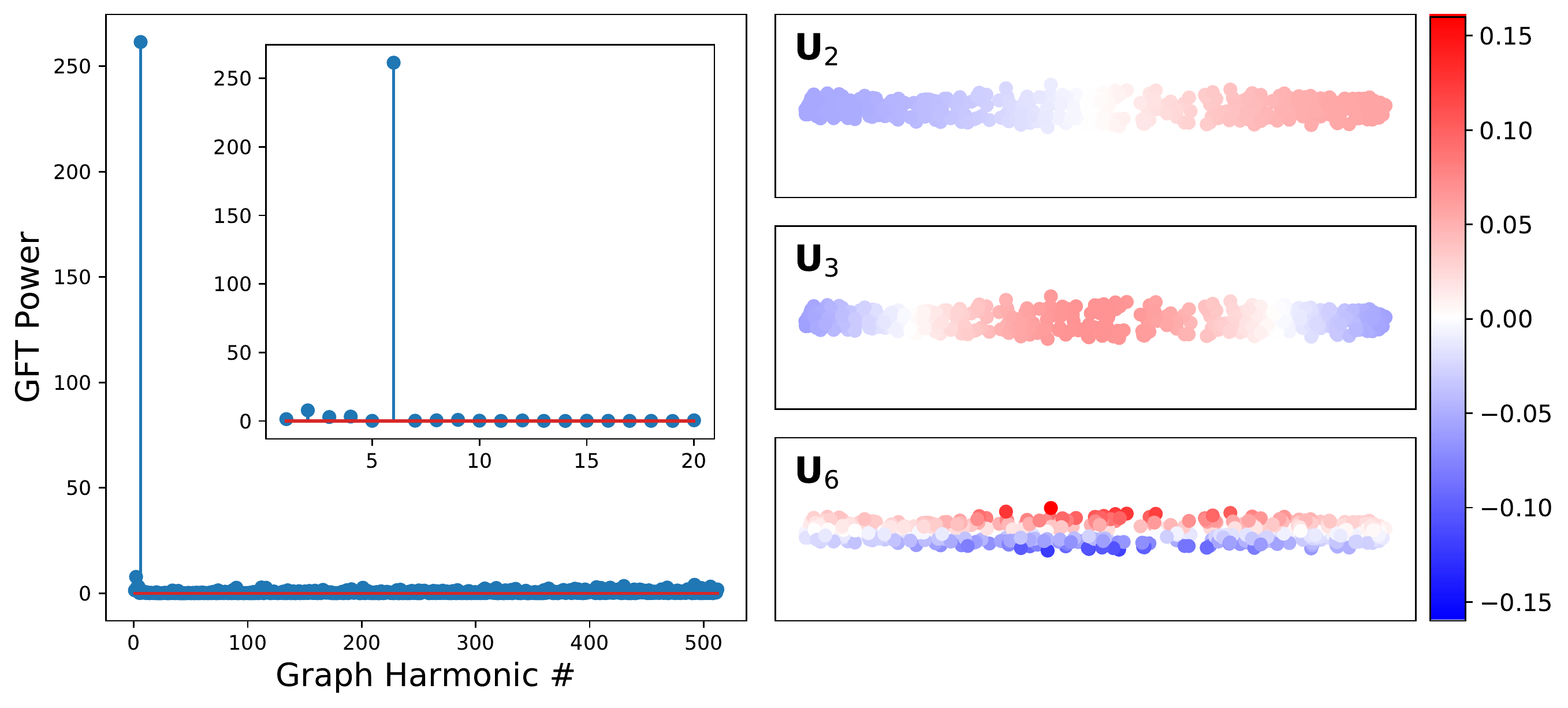}

    \caption{GFT Power of the signal $\mathbf{u}$ and sample harmonics for a flattened ring-like shape using a squared exponential weight and the normalized Laplacian. When applied to the graph signal $\mathbf{r}$, the signal is approximately $99\%$ concentrated in $\mathbf{U}_2$. Tangent motion along this curve has even less angular momentum than above, $m_a\approx0.18$}
    \label{fig:flat_GFT_exp}

\end{figure}

To further pursue this analysis, consider the a notional state using the curve $(x(t),y(t))=(3\cos(t)+0.1\sin(t))$, and otherwise defined as above.  This produces a ``flattened'' ring state (see Fig.~\ref{fig:flat_GFT_exp}) and is reminiscent of the state in \cite[Fig.~5b]{strombom2011collective}. In this case, the topological hole in the swarm has essentially vanished. In this case, we found that the disk-based adjacency matrix did not have a particularly structured spectral response for $\mathbf{u}$ using either Laplacian. However, the weighted exponential using the normalized Laplacian placed roughly half the \ac{GFT} power into a single harmonic and was almost evenly distributed outside of this.  Additionally, all four combinations of graph and \ac{GFT} resulted in a single dominant ($>97\%$ signal power) in the second harmonic when $\mathbf{r}$ was used as the graph signal.  This should not be particularly surprising when one observes that the swarm itself is essentially one dimensional in this configuration.  Thus, in this case we have that the \ac{GFT}s of $\mathbf{u}$ and $\mathbf{r}$ barely overlap at all, and furthermore there is little angular momentum $m_a\approx0.18$.

\subsection{Curve States}


Building on the analysis in the previous section, consider a notional state derived from the Lissajous curve $(x(t),y(t))=(3\sin(t+\frac{\pi}{4}),\sin(3t))$, and the velocity defined analogously to the examples in the previous section. This produces a curve that is irregularly shaped and self-intersecting, see Fig.~\ref{fig:curve_state_GFT_harms}.
States similar to this can form as highly polarized swarms make sharp changes of direction (c.f., Figure~6 in {\cite{tunstrom2013collective}), by directed motion that intersects itself \cite[Fig.~5]{strombom2011collective}, or by phase sorting in mobile coupled oscillators  \cite[Fig. 3]{monaco2020cognitive}.
This particular state has negligible angular momentum due to its particular symmetries.






\begin{figure}[!ht]

  \centering
  \includegraphics[width=\columnwidth]{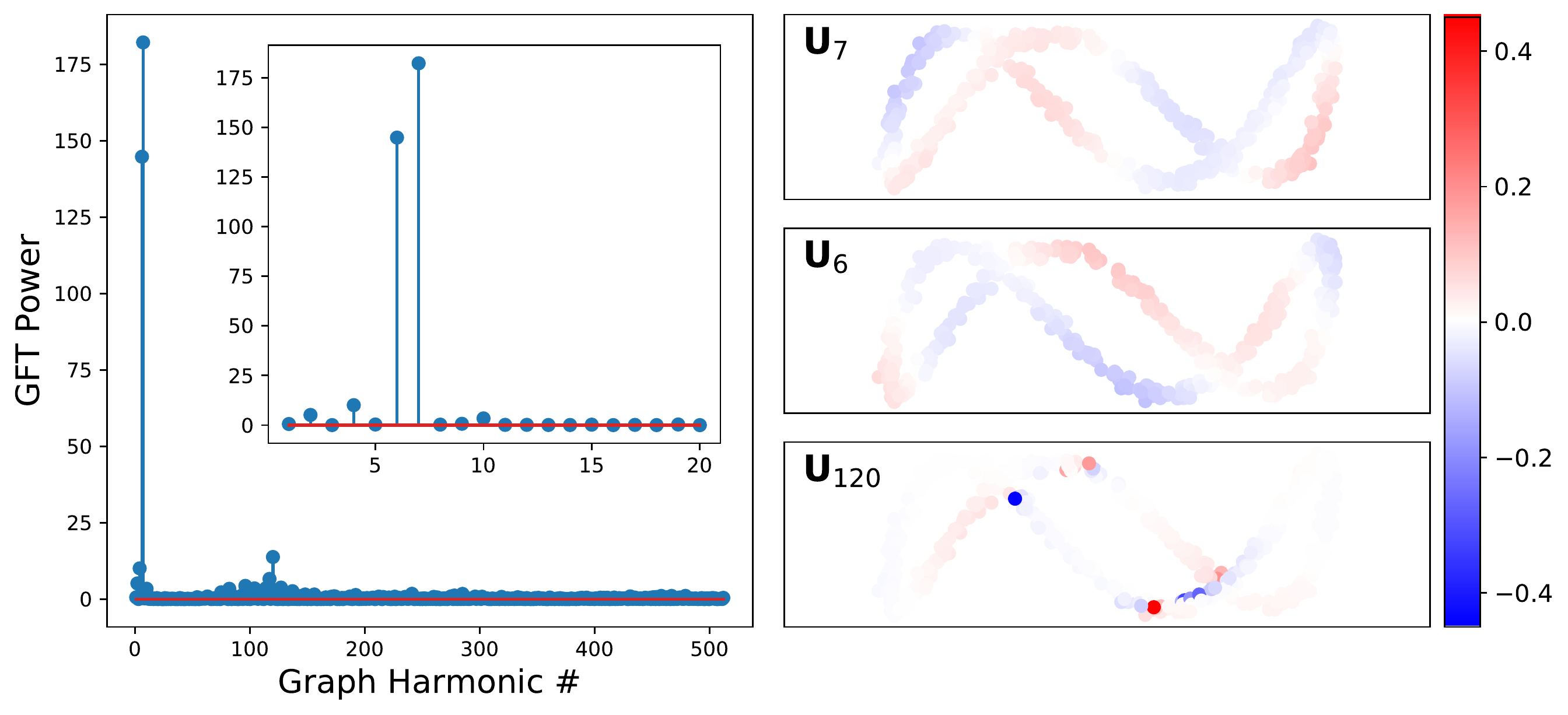}

  \caption{\ac{GFT} power of $\mathbf{u}$ and top three harmonics derived using the normalized Laplacian of a squared-exponentially weighted graph.}

  \label{fig:curve_state_GFT_harms}

\end{figure}

As before, we applied all four combinations of disk-based vs.\ weighted graphs and both forms of the Laplacian.  All four combinations had strong power concentration in $\mathbf{U}_2$ which divides the notional state along the line $x=0$, and secondary concentration in harmonics that divide the notional state along the line $y=0$ (for an example see \cite{supp}).  When $\mathbf{u}$ is considered in the graph Fourier domain, as was the case with the flattened ring, we found that the weighted graph with the normalized responses produces by far the most structured and with power concentrated primarily into two higher order harmonics.
The third highest spectral peak occurs at by far the highest graph frequency of any notional example considered thus far, but itself has notable outlying values at the self-intersection points. 
An open question here is that if a mechanism for generating a directed graph that resolved the self-intersection issue could be used to define a \ac{GFT} that would have harmonics that track the two dimensions of the velocity more closely.






To further illustrate the flexibility of this approach, consider a notional swarm state that is defined on an open curve (specifically, the same Lissajous curve as above for $t\in[0,\pi)$), with velocity $\mathbf{v}$ tangent to the curve as before (see Fig.~\ref{fig:curve_state_open_harms}).  Unlike the other states considered up to this point, this state should have a substantial portion of the \ac{GFT} power of $\mathbf{u}$ in the first harmonic, as there is a net imbalance of $\mathbf{u}$ along the portion of the curve considered.  Traditional order-parametric approaches to swarm analysis (as in \cite{tunstrom2013collective}) would first look at the quantity $||\bar{\mathbf{u}}||_2~\approx 0.38$, which could, for example, be consistent with a somewhat disorganized ``cloud'' that is heading in a consistent direction.  Next, they might consider $m_a\approx 0.254$.  Such parameters would be consistent with a swarm in a mixed state between a polarized and milling behavior, for example.
Using the \ac{GFT} analysis, however, we see that there is considerable power in a contiguous band of graph harmonics, i.e., there is a very specific bandlimited structure of $\mathbf{u}$ (see Fig.~\ref{fig:curve_state_open_harms}). 

\begin{figure}[!ht]

  \centering
  \includegraphics[width=\columnwidth]{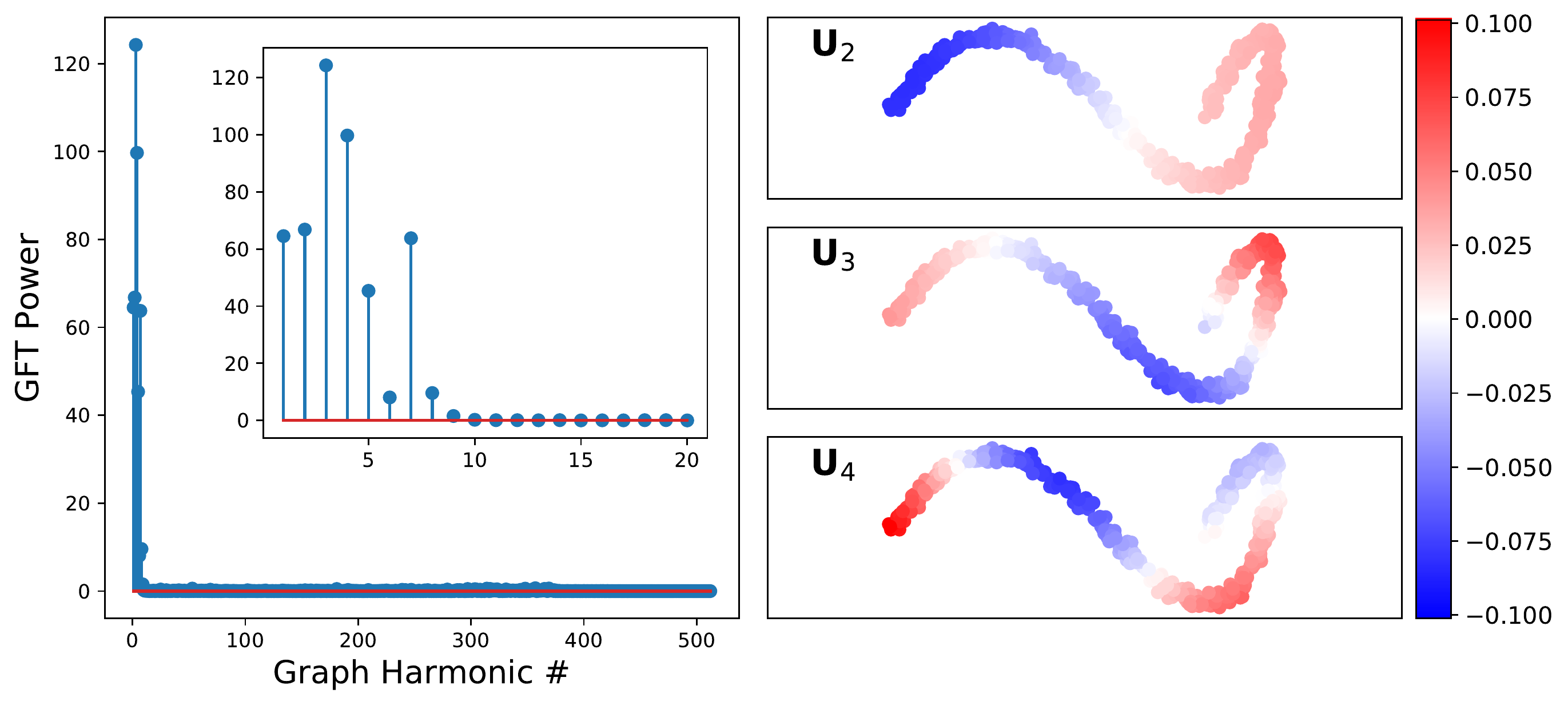}

  \caption{\ac{GFT} power of $\mathbf{u}$ and top three harmonics derived using the normalized Laplacian of a squared-exponentially weighted graph. When $\mathbf{r}$ is used as the graph signal, the concentration of power in harmonics 2-4 are $\approx86\%$, $8\%$, and $4\%$, respectively.}

  \label{fig:curve_state_open_harms}

\end{figure}

In total, the combination of these different graph and \ac{GFT} approaches to these less symmetric rings  and curve states demonstrate that \ac{GSP} 
is capable of revealing structure in ways that standard order parameters are not.  Furthermore, as swarms become more complex and irregular, it would appear that there may be value in considering weighted graphs using the normalized Laplacian, as this combination appears to further refine the geometric structure of the swarm.


\section{Simulated Swarm States}
In this section, we apply the \ac{GSP} analysis considered in the previous section to a series of simulated swarming models, both demonstrating the utility of the techniques to actual dynamical swarming models and providing additional context to discuss design considerations in using \ac{GSP} to analyze swarms. In particular, we highlight:

\begin{itemize}
  \item (Example 1: Vicsek-Style Swarm \cite{costanzo2018spontaneous}) the natural extensibility of the \ac{GSP} approach in analyzing swarms that become fractionated into multiple sub-swarming components,
  \item (Example 2: Couzin et al \cite{couzin2002collective}) the impact of choice of graph definition and \ac{GFT} in analyzing diffuse swarms, 
  \item (Example 3: Swarmalators \cite{o2017oscillators}) the ability of \ac{GSP} approaches to analyze graph signals beyond velocity and graphs beyond distance-based adjacencies.
\end{itemize}

\subsection{Example 1: A Vicsek-Style Swarm}

A classic model of swarming in self-propelled particle systems and active matter is the Vicsek model \cite{vicsek1995novel} and its many generalizations, e.g.,\cite{huepe2004intermittency,gregoire2004onset,aldana2007phase,chate2008modeling,costanzo2018spontaneous}.  Agents in these models generally consist of a pair of states $\mathbf{x}_j$ and $\mathbf{v}_j$ representing the position and velocity, respectively. In the original formulation, and many of its successors, a fixed magnitude $\mathbf{v}_j$ is assumed, resulting in velocity states represented by angles $\phi_j$ (equivalently, unit vectors $\mathbf{u}_j$ for 2 or more dimensions). In the original Vicsek model, the dynamics of the velocity state included only velocity alignment to agents nearby in a fixed sensing radius \cite{vicsek1995novel}, but its various extensions have introduced models exhibiting alignment driven primarily by positional repulsion \cite{grossman2008emergence}, positional attraction \cite{strombom2011collective}, and both \cite{gregoire2003moving}. Additional modifications such as restricted fields of view and turning rates 
\cite{costanzo2018spontaneous} or long-range interactions \cite{kruk2018self} can result in additional collective behaviors.

In this section, we use the \ac{GFT} analysis on a particularly striking simulation run using the modified Vicsek model from \cite{costanzo2018spontaneous}.  The primary difference between this model and the standard Vicsek model is the introduction of a limited field of view, where the alignment neighborhood is missing a blindspot opposite of the direction of agent motion, and a hard limit on the maximum angular velocity of the agents.  This results in ``milling'' behavior, where there is directed motion along an approximately annular swarm structure. Furthermore, there may be multiple mills and/or coherently aligned components in the overall swarm.

To first consider the simulation run from a purely topological perspective, we use the simulation agents' sensing range as a hard cut-off to define a symmetric adjacency matrix, ignoring the blind spot.  Using the dimension of the null space of the corresponding combinatorial Laplacian, we can compute the number of connected components in the swarm as a function of time, shown in the top panel of Fig.~\ref{fig:vicsek_comps}.  First we note that there is an initial phase of expansion and contraction as the swarm eventually settles into five connected time components where it remains for some 1000 time steps before one of the components dissolves resulting in several transient components before settling into four milling states. These milling components remain stable until at least 10000 time steps (not shown).

\begin{figure}[ht!]

\includegraphics[width=\columnwidth]{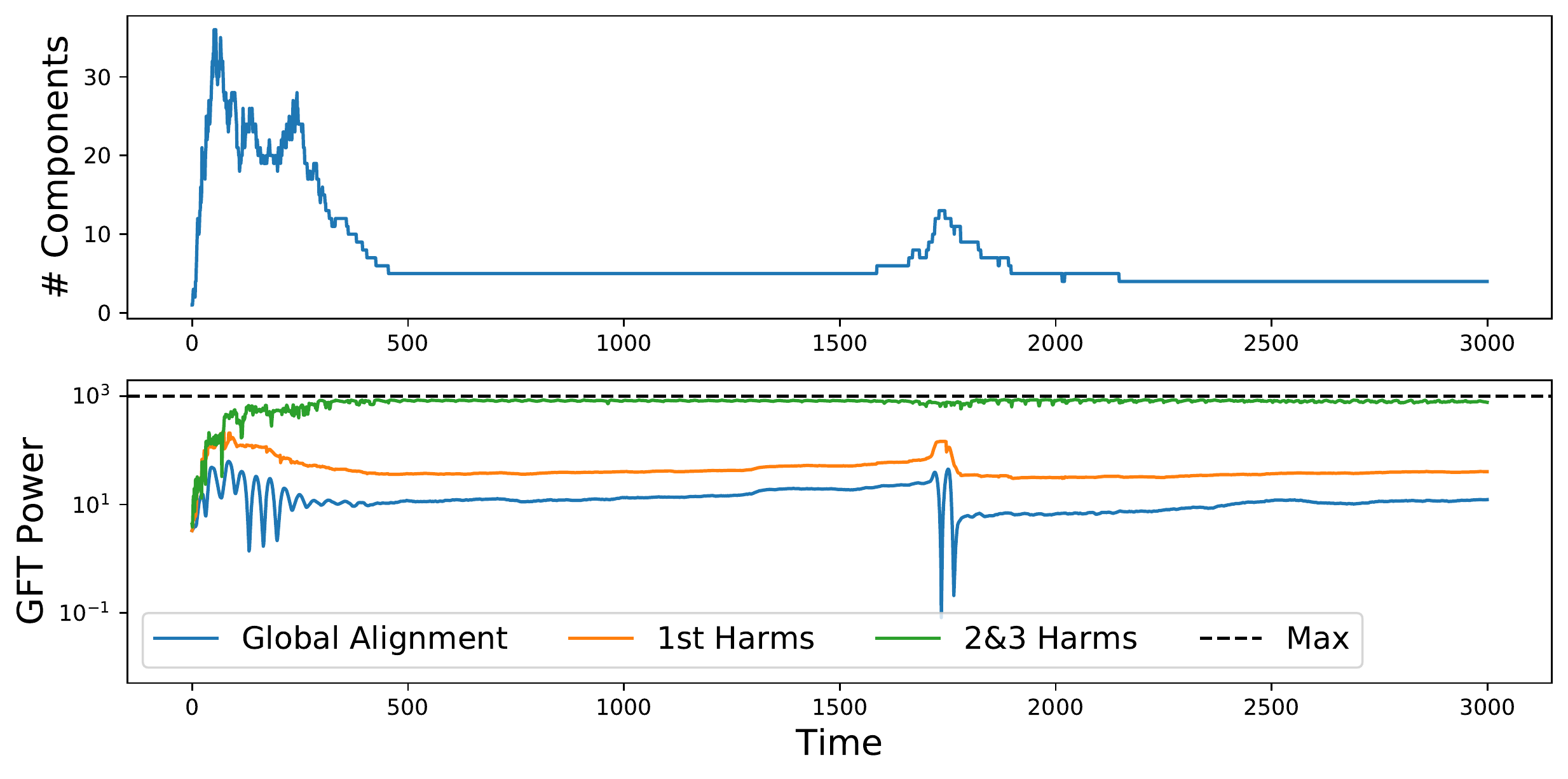}

\caption{Top: Number of connected components in a sample simulated swarm over time, using the agents' visibility cutoff (1 unit) to define a graph topology. Bottom: Plots of \ac{GFT} power over time in 1) a global alignment harmonic, 2) collective power in each connected component's first harmonic (i.e., alignment within a connected component), and 3) collective power in each connected component's second and third harmonic.  The swarm state depicted in Fig.~\ref{fig:vicsek_five} corresponds to time step 1500, shortly before the upper left milling component unravels, dissolves, and is absorbed into the remaining milling components.}

\label{fig:vicsek_comps}

\end{figure}

To illustrate how the \ac{GSP} approach to swarm analysis naturally extends to the complex trajectory considered here, Fig.~\ref{fig:vicsek_five} shows the swarm state and corresponding \ac{GFT} of the swarm at time step 1500, when there are five connected components. Using our definition of the Laplacian for disconnected graphs, Fig.~\ref{fig:vicsek_five} shows the second graph harmonic for each connected component, demonstrating that the intuition from the ring and annular states from the previous section should still hold here. As we expect, given that agents in the milling states have headings tangent to the ring structure, the \ac{GFT} of the graph signal $\exp(i\phi_j)$ shows strong concentration in the second and third harmonics of each connected component.  Additionally, we note that the two components at the top of the panel appear to be both the least evenly distributed across their mills (possibly due to their smaller size), and also appear to be dominated by the second harmonic, as compared to the other three that have a more even distribution in power between these two harmonics.

\begin{figure}[ht!]

\includegraphics[width=\columnwidth]{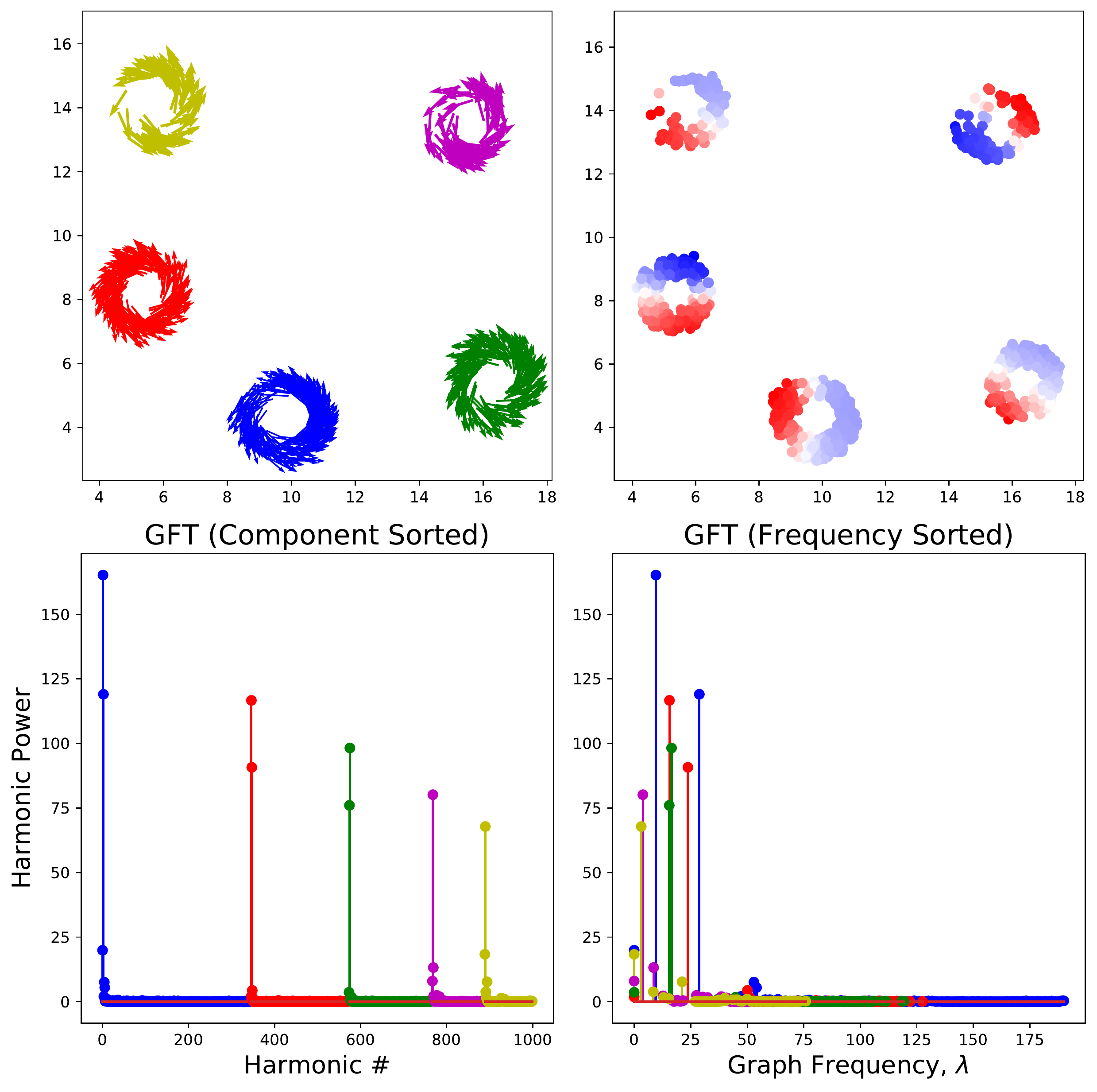}

\caption{(Color Online) Example \ac{GFT} decomposition of a complex swarm. Upper Left: A multi-component swarm using the model of \cite{costanzo2018spontaneous} showing five disconnected milling behaviors (rotating annuli).  The quiver plot arrows show the direction of motion for each agent. Each component is a different color for illustration purposes. Upper Right: Depiction of the second graph harmonic for each connected component. For presentation purposes, the scale for each component is different due to the differences in component size but blue represent negative weights and red positive as is the convention throughout this paper. Bottom Left: \ac{GFT} decomposition of the swarm color coded by component, ordered by decreasing component size and graph frequency. Bottom Right: \ac{GFT} decomposition of the swarm color coded by component, ordered by increasing graph frequency. Note each individual mill is dominated by a pair of harmonics (corresponding to the first two non-zero eigenvalues from each component's Laplacian).}

\label{fig:vicsek_five}

\end{figure}

We also note that examples such as this point out the difference between considering the power distribution ordinally (bottom left) as compared to solely a function of graph frequency (i.e., eigenvalue of the Laplacian).  While the frequency-based viewpoint demonstrates an overall sparse and low frequency graph signal, the ordinal viewpoint reinforces the notion that the graph signal power is concentrated in harmonics in an intuitive manner.  Furthermore, since the frequency of a given harmonic is highly dependent on the specific connectivity (and component size for the combinatorial Laplacian), the graph frequencies will not be as stable as their sorted order (especially within a connected component).  This allows for a simpler comparison between swarm states over when considering only the ordinal value of harmonics. There are, however, \ac{GSP} techniques for dynamic (i.e., time-varying) graphs that evaluate a time series of graph signals along common (with respect to time) \ac{GFT} harmonic subspaces, but we leave these computationally intensive techniques for future work.  Finally, by design, we expected that the swarms considered here would have sparse, low frequency representations, but other examples of collective motion may exhibit power law trends (c.f.,  \cite{expert2017graph}), in which case considering the actual graph frequency would be required.

The bottom panel of Fig.~\ref{fig:vicsek_comps} shows aggregated analysis of the entire swarm trajectory.  As one might expect, the global alignment measure $|\frac{1}{\sqrt{N}}\sum_j\exp(i\theta_j)|^2$ (i.e., the first graph harmonic if the swarm were connected), is quite small throughout the simulation run.  Similarly, the local alignment measured by the first harmonic of the individual connected components is also quite low, but due to imbalances in the milling states is slightly higher than the global alignment. Finally, as the majority of the swarm agents are in milling states the majority of the simulation run, we see that the sum of the powers in the second and third harmonics accounts for nearly all of the potential \ac{GFT} power.

We stress that the analysis here can be completely automated, if, for example the data was generated from observations of experiment of e.g., fish. The only stage at which we have specified a parameter is in the assumed sensing radius of the swarming agents.  Equivalent analysis could be produced by applying \ac{TDA} persistent homology techniques as in \cite{topaz2015topological,corcoran2017modelling,sinhuber2017phase} to find a sensing radius that results in large persistence range for a few connected components. Alternatively, one could use a decaying kernel to define a weighted adjacency matrix to produce similar results \cite{supp}.







\subsection{Example 2: Couzin et al}
In \cite{couzin2002collective} a model for swarming in three dimensions was introduced that relied on discrete ranges of interaction for repulsion, alignment, and attraction terms, as well as a conical ``blind-spot'' behind the swarming agents. There, it was demonstrated that varying the relative ranges of the three interactions produces four fundamentally different regimes.  The first, that they denoted the ``swarming '' regime is characterized by a lack of alignment on both  the global scale, resulting in little motion of the swarm center of mass, as well as a lack of alignment locally, resulting in little angular or absolute angular momentum.  As the radius of alignment increases, the ``torus'' state appears, characterized by a high level of angular momentum.  As the radius of alignment continues to increase, they denote the collective behavior ``dynamic parallel'', where the global alignment of the swarm increases and the swarm as a whole travels from its starting point.  In this state, the swarm has some variability in velocity and the individual noise fluctuations on the agent headings are readily apparent.  As the radius of alignment approaches the radius of attraction, the swarm dynamics produce ``highly parallel'' states, where the headings are essentially uniform and the swarm travels at a velocity approaching the individual agents' velocities.

Unlike the model in the previous section, this swarming usually produces connected swarms, but these tend to be more diffuse and irregular (at least partially due to the three-dimensional nature of this model).  Additionally, we find that the inter-agent distances varies as a function of the equilibrium state, with swarm and torus states considerably more diffuse than either the dynamic parallel or highly parallel states. This irregularity, combined with the potential for long-term transient behavior in the model motivates the desire to have an adaptive model for determining the topology of the network in a way that accounts for the variable density over time.  In principle, we could apply the persistence based techniques from \cite{topaz2015topological,sinhuber2017phase}, but as we say in the notional swarming section, it may be beneficial to use a graph that weights the edges as a decaying function of distance. One with this approach is the fact that the different behavior regimes in \cite{couzin2002collective} have different spatial scales, so instead we use the (normalized)

Ultimately, the aim of \ac{GSP} is to exploit structure in the \ac{GFT} domain generally in the form of bandlimitedness or sparsity \cite{sandryhaila2013discrete,shuman2013emerging,ramakrishna2020user}.  Anecdotally, we observed that the normalized Laplacian using either the unweighted or weighted adjacency matrix appeared to produce sparser signals in the graph Fourier domain (in addition to producing more continuous transformed signals as function of time) \cite{supp}. To formalize this intuition we performed 20 Monte Carlo runs of 100 swarming agents for each of the parameter regimes in \cite{couzin2002collective} and used a metric from the compressive sensing literature, the Gini sparsity metric \cite{hurley2009comparing} to measure the resulting sparsities of the final 750 time steps of simulations lasting 1500 time steps.  Fig.~\ref{fig:couzin_gini_comb} shows the distributions of these sparsity values as violin plots, which indicate that the normalized Laplacian approaches, $\bar{\mathbf{L}}$ and $\bar{\mathbf{L}}_{exp}$ have improved sparsity over their combinatorial Laplacian counterparts, and that furthermore the combination of the normalized Laplacian with the exponentially weighted adjacency matrix has the best overall sparsity across the four behavior regimes.  This suggests that for the \textit{application} of \ac{GSP} techniques that seek to exploit structure, the the transform defined by $\bar{\mathbf{L}}_{exp}$ will perhaps offer the most utility.

\begin{figure}[!ht]

  \centering
  \includegraphics[width=\columnwidth]{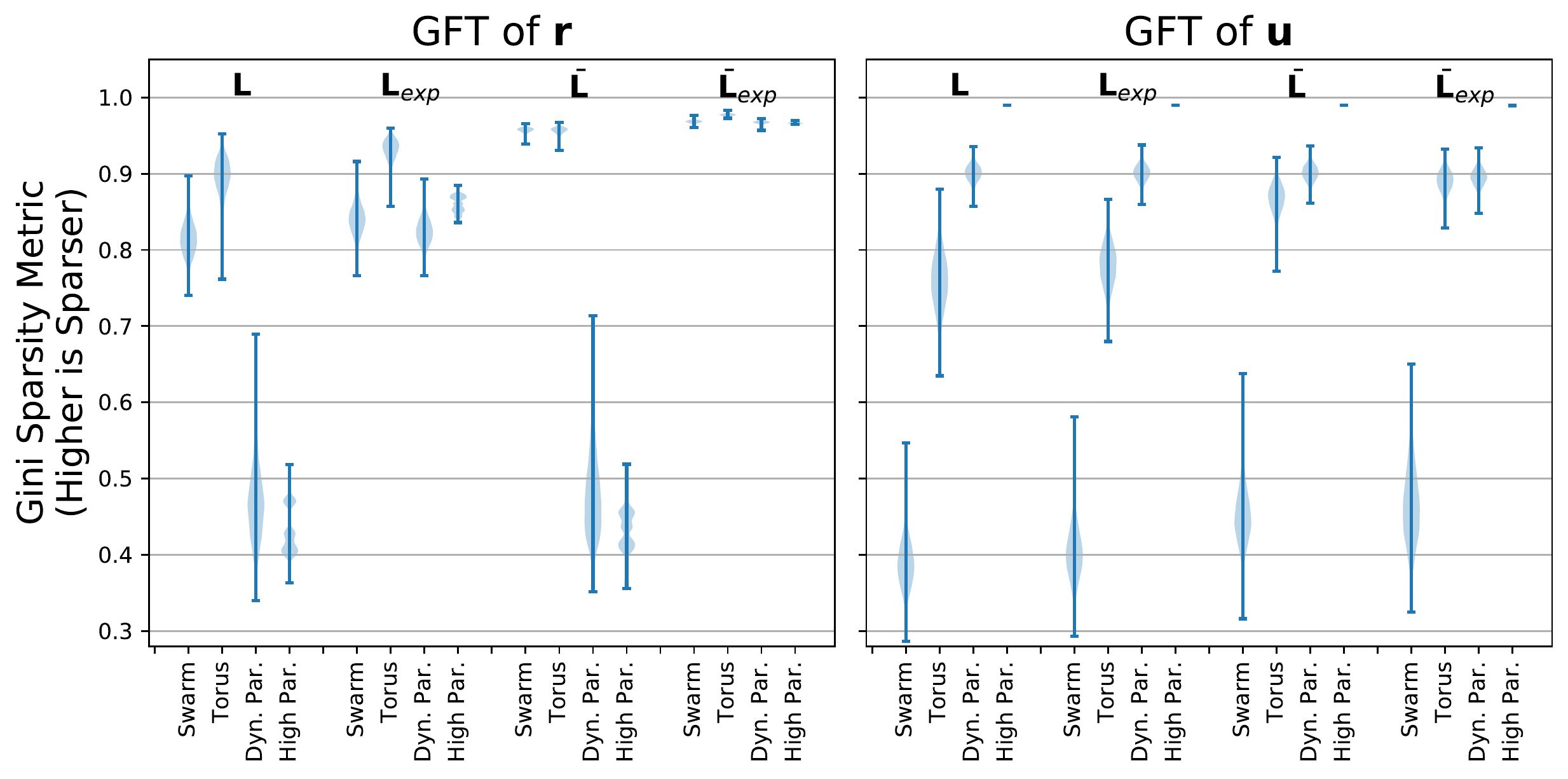}

  \caption{Comparison of \ac{GFT}-induced signal sparsity.  Violin plots of the Gini sparsity metric \cite{hurley2009comparing} using different \ac{GFT} methods applied to the final 750 time steps of a 1500 step run for 20 random simulation runs. Left: Sparsity of the spatial signal $\mathbf{r}$. Right: Sparsity of the velocity signal $\mathbf{u}$.  The normalized Laplacian based on the squared exponential kernel produces the sparsest signals for both $\mathbf{r}$, and $\mathbf{u}$ (albeit generally marginally for $\mathbf{u}$).  Since the ultimate goal of \ac{GSP} is to exploit spectral structure, this demonstrates that the choice of transform can drastically influence this structure.}

  \label{fig:couzin_gini_comb}

\end{figure}

Despite these differences in transformed signal sparsity for these different transforms, we see that their total variations are generally consistent (up to an overall scale) for both the graph functions $\mathbf{r}$ (Fig.~\ref{fig:couzin_r_tv}) and $\mathbf{u}$ (Fig.~\ref{fig:couzin_u_tv}).  This indicates that improved sparsity (i.e., more exploitable structure) in the \ac{GFT} signals is does not appear to be at the cost of losing some more macro-scale signatures.  In particular, we see that the general trends for all four combinations of graph and transform method produce the same oscillatory or damped transients (depending on the swarm behavior) and the same relative orderings of the four spatial states. The exception to this latter point appears to be the transform from $\bar{\mathbf{L}}_{exp}$, which provides better discrimination (i.e., separation in total variation) between the swarming and torus behaviors, further reinforcing the utility of this particular transform for the model considered here.  These results would appear to serve as the basis for a robust classifier of these different swarm behaviors as in \cite{berger2016classifying} and it appears that there are several long-running transients in the dynamic parallel regime.

\begin{figure}[!ht]

  \centering
  \includegraphics[width=\columnwidth]{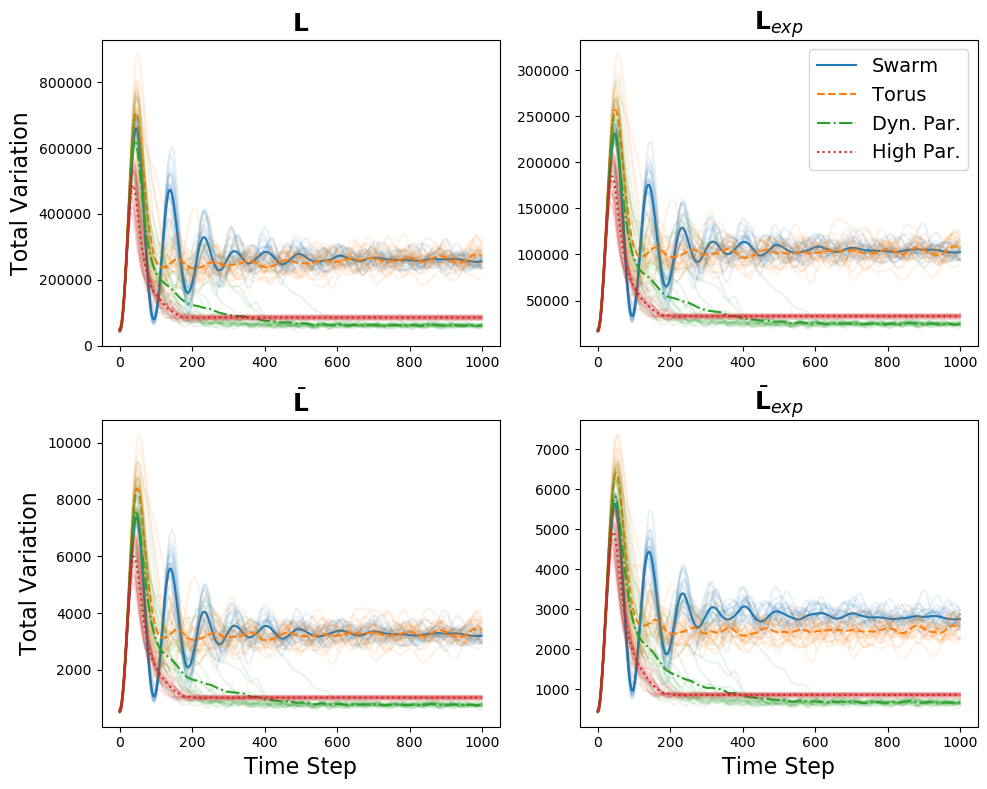}

  \caption{Plots of \ac{TV} of $\mathbf{r}$ over time using different \ac{GFT}s. The translucent lines are the individual Monte Carlo runs of the simulation in \cite{couzin2002collective}, and the darker lines the respective means for the four different swarm states.}

  \label{fig:couzin_r_tv}

\end{figure}

\begin{figure}[!ht]

  \centering
  \includegraphics[width=\columnwidth]{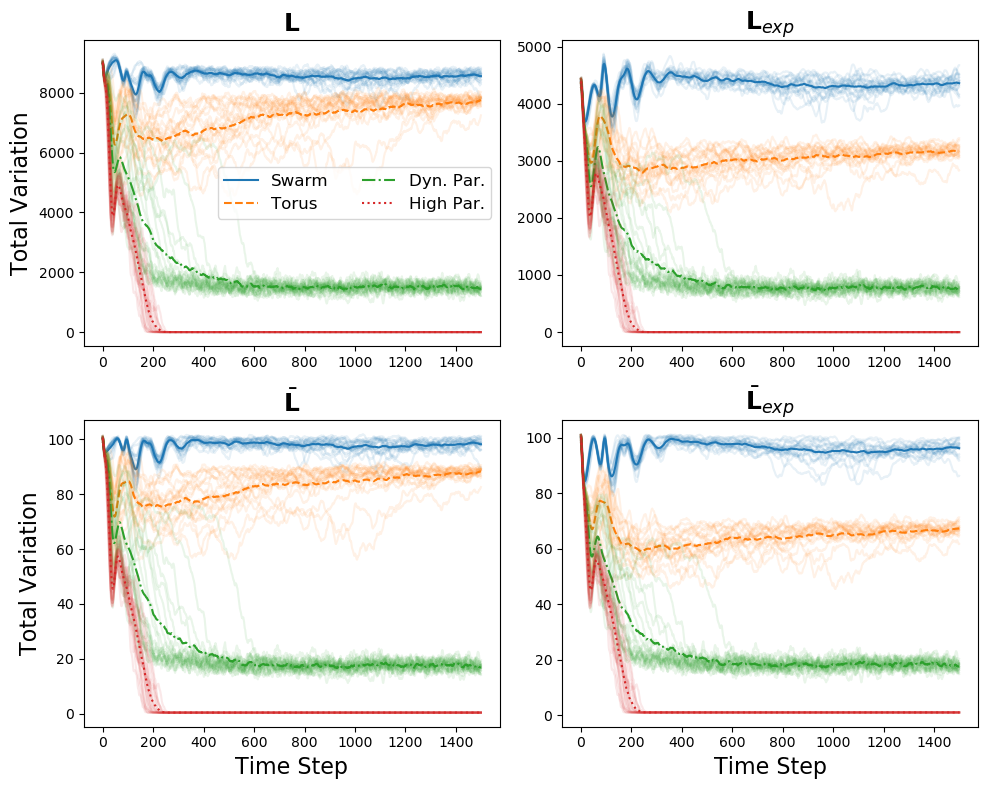}

  \caption{Plots of \ac{TV} of $\mathbf{u}$ over time using different \ac{GFT}s. The translucent lines are the individual Monte Carlo runs of the simulation in \cite{couzin2002collective}, and the darker lines the respective means for the four different swarm states.}

  \label{fig:couzin_u_tv}

\end{figure}

\subsection{Example 3: Swarmalators}
In contrast with Vicsek style models that often incorporate an angular variable $\phi_j$ and study positional alignment, the swarmalator model \cite{o2017oscillators} combines spatial states $\mathbf{x}_j$ with an angular state $\theta_j$ and studies interactions between spatial aggregation (i.e., swarming) and synchronization in oscillators.  Several formulations of these dynamics were explored in \cite{o2017oscillators} as well as other extensions in \cite{hong2018active,o2018ring,o2019review,jimenez2020oscillatory}. A simple form of these dynamics from \cite{o2017oscillators} are: 
\begin{equation}
\begin{aligned}
\dot{\mathbf{x}}_j&=\frac{1}{N}\Biggl[\sum_{k\neq j}^N \frac{\mathbf{x}_k-\mathbf{x}_j}{||\mathbf{x}_k-\mathbf{x}_j||_2}(A+J\cos(\theta_k-\theta_j))\\
& \hspace{3.5cm}- B\frac{\mathbf{x}_k-\mathbf{x}_j}{||\mathbf{x}_k-\mathbf{x}_j||_2^2}\Biggr]\\
\dot{\theta}_j&= \frac{K}{N}\sum_{k\neq j}^N\frac{\sin(\theta_k-\theta_j)}{||\mathbf{x}_k-\mathbf{x}_j||_2}
\end{aligned}
\end{equation}
where $A$, $B$, $J$, $K$, are scalars that determine fundamentally different behavior regimes. 

In \cite{o2017oscillators} several different phase transitions were observed for this formulation of swarmalator dynamics. In particular, for the case $A=B=J=1$, and $K$ sweeping from $0$ to $-1$ several fundamentally unique steady states are observed (see Fig.~\ref{fig:swarmalator_steady_states}). When $K=0$, there are no phase dynamics, and the swarmalators align in a phase-sorted annular state. For small negative $K$, the interplay between phase alignment and positional repulsion from dissimilar phases results in a splintered wave state that produces a series of clustered ``wedges'' along an annular structure.  As $K$ continues to decrease, the splintered wave gives way to an active wave state, with the swarmalator agents changing in phase and spatially along a roughly annular structure.  Eventually, the ``hole'' of the annulus is not consistently present due to the mixing of the agents spatially.

\begin{figure}[!ht]

  \centering
  \includegraphics[width=\columnwidth]{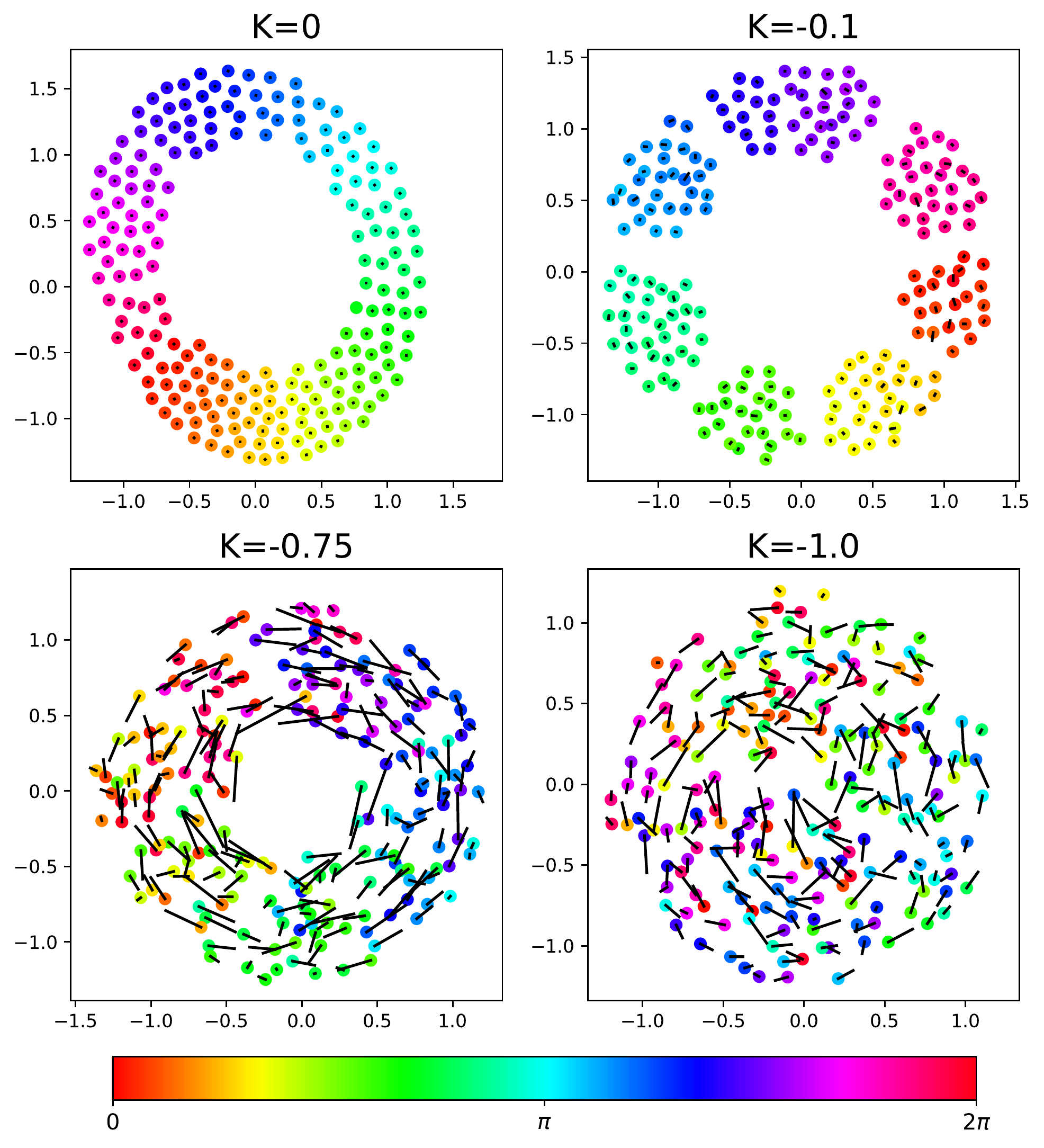}

  \caption{Sample swarmalator states for $A=B=J=1$ and various $K$.  Color indicates the swarmalator phase state $\theta_j$, and the line segments indicate velocity. A $K$ decreases, the swarmalator steady states transition from the static phase wave state, to the splintered wave state, to active wave states.}

  \label{fig:swarmalator_steady_states}

\end{figure}

The swarmalator model presents an opportunity to highlight some additional use cases of \ac{GSP} analysis beyond the general process already considered.  In particular, the swarmalator model introduces another angular variable to study, $\theta_j$, beyond the velocity heading $\phi_j$ that we have already considered.  Additionally, the state $\theta_j$ can be used as a different, non-Euclidean coordinate to define a graph. First, we consider the graph defined using $\exp(||x_j-x_k||_2^2/\sigma^2)$ where $\sigma^2$ is the average inter-agent distance squared. In \cite{o2017oscillators}, the correlation between phase $\theta_j$ and position was evaluated by considering the maximum of $|\frac{1}{N}\sum_{j}\exp(\theta_j\pm\psi_j)|$, a computation equivalent to computing angular momentum in both the clockwise and counter-clockwise directions.  They found that this alignment tracked the value of $K$.  Given our results from previous sections, we would expect to see this reflected as a concentration of \ac{GFT} power in the second and third graph harmonics of the respective swarmalator states.

To verify this, for each $K$ between $0$ and $-1$ (inclusive) in increments of $0.05$, we generated 20 random swarmalator instances and ran the evolutions 5000 time steps. Using $\theta_j$ as a graph function for \ac{GSP} analysis using the normalized Laplacian yields graph signals with a substantial portion of their overall power in the first and second harmonics (see Fig.~\ref{fig:swarmalator_power_violins} (left)).  This generally correlates with $K$, due to the emergence of active phase wave state, but there are some finite size effects for $|K|$ small. Using the same connectivity, we can perform similar analysis using the agents' headings, $\phi_j$, as a graph signal.  Unlike the other swarm models considered in previous sections, swarmalators do not exhibit concentration in the second and third harmonics, and are ``white'' graph signals in general.  We do note, however, that the analysis for the $K=0$ state is omitted as the velocities are basically zero in magnitude but the discrete time implementation we use here introduces random fluctuations in the sign of the velocity, leading to unstable results.

\begin{figure}[!ht]

  \centering
  \includegraphics[width=\columnwidth]{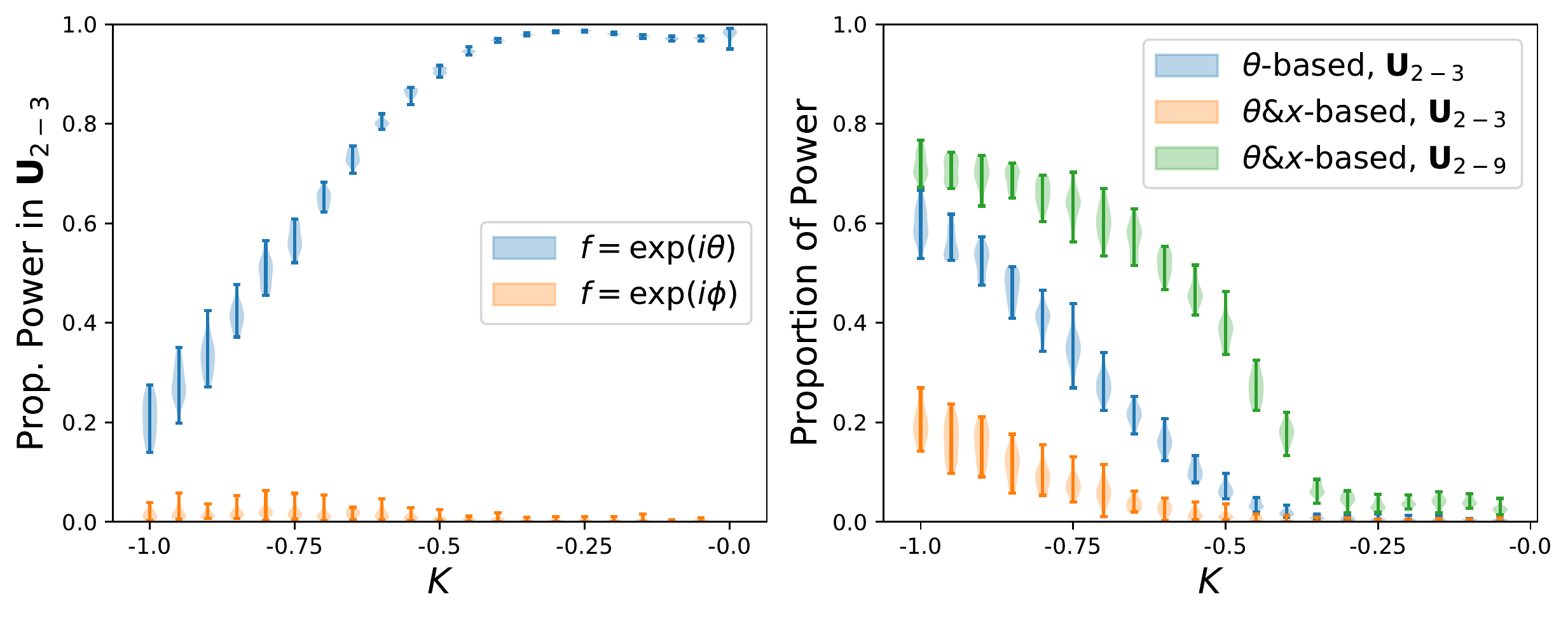}

  \caption{Violin plots of swarmalator \ac{GFT} power. Left: Proportion of signal power using the in the second and third harmonics using the graph derived from $\mathbf{x}$, for graph signals based on $\theta_j$ (swarmalator phase state state) and $\phi_j$ (swarmalator velocity heading). Right: Proportions of \ac{GFT} power in various bands (see legend) using graphs based on just the distances between phase states $\theta_j$ and using both phase and spatial states.}

  \label{fig:swarmalator_power_violins}

\end{figure}

The fact that the position-based graph does not uncover any usable structure in the heading signal suggests that we should construct alternative graphs to elucidate any relationships.  In particular, one might conjecture from the plots in Fig.~\ref{fig:swarmalator_steady_states} there is some alignment in heading as a function of difference between the phase variables.  To investigate this, we use angular distance $d(\theta_j,\theta_k)=\arccos(\cos(\theta_k-\theta_k))$ to construct a graph $A_{jk}=\exp(-d(\theta_j,\theta_k)^2)$, $A_{jj}=0$ and corresponding \ac{GFT} using the normalized Laplacian. Fig.~\ref{fig:swarmalator_power_violins} (right) shows how the proportion of signal power of the agent headings in the second and third harmonics increases as $K$ decreases and enters active phase wave states. For values closer to zero, the signal is unstructured in the \ac{GFT} domain, indicating little connection between smoothness in phase and heading. 

We further pursue this line of inquiry by combining the spatial and phase distances to construct a graph that considers jointly the spatial and phase relationships.  Specifically, we use $A_{ij}=\exp\left(-\left(||x_j-x_k||_2^2+d(\theta_j,\theta_k)^2\right)/\sigma_{tot}^2\right)$ where $\sigma^2_{tot}$ is the average combined sum of the inter-agent spatial and phase squared-distances, with $A_{jj}=0$ and again consider the normalized Laplacian. In this case (see Fig.~\ref{fig:swarmalator_power_violins}), this graph has a small amount of power concentration in the second and third harmonics as $K$ decreases, much less so than the \ac{GFT} using the phase distance only.  However, this set of transforms yields an obvious low pass signature resulting in spectral concentration in harmonics 2-9 that is not observed using either the distance-only or phase-only graphs. 
For values of $K\geq-0.45$ we find that this concentration accumulates in harmonics 8 and 9, and starting at $K=-0.5$, this concentration switches to harmonics 6 and 7.  As harmonics 2 and 3 start to contribute, we actually see that all harmonics between  2 and 7 are contributing to the signal concentration. 


%
\section{Discussion and Future Directions}
In conclusion, we have shown how swarms can be embedded into natural graphical structures in the vein of computational topology and \ac{TDA}, and that swarm states defined on these graphs can be decomposed into Fourier harmonics that respect natural geometric structure implied by these graphs.  This graph Fourier analysis reveals that many common swarming behaviors result in highly structured signals when viewed in the graph Fourier domain.  The work presented here represents a broad, but shallow, cut through a wide range of notional and simulated swarming states and we believe there is considerable future research in investigating specific swarming models as well as experimental data using these concepts and insight.  The discussion of the nuances of the choice of graph, graph function, and the different forms of \ac{GFT} will serve as a valuable resource for these future endeavors.

\ac{GSP} is fundamentally about the generalization of signal processing techniques, and when graph signals exhibit structure in the graph Fourier domain many techniques exist to solve problems in inference and signal conditioning (for a recent review see \cite{ramakrishna2020user}).  Such potential applications of \ac{GSP} to the analysis of collective motion include filtering to denoise noisy data, graph Fourier-based clustering for unsupervised learning of both collective and individual behaviors,  and estimation of global swarm states from sparse measurements.  Already, we have applied the techniques from this paper to the detection of anomalous agents in an otherwise nominal swarm \cite{schultz2021detecting}.  Beyond such concrete applications of \ac{GSP}, the analysis in this work focused primarily on graph frequency in an ordinal sense, but similar investigations of other swarming behaviors may reveal power-law type trends when the frequencies are considered in an absolute sense, similar to the analysis of \cite{expert2017graph}.

There are other signal transform techniques from the field of \ac{GSP} that could be brought to bear in the analysis of collective motion.  Here, we essentially considered each time step of the swarm as both a separate graph and graph signal.  However, \ac{GSP} naturally extends to transforms in both the time and vertex domains \cite{shuman2016vertex,grassi2017time}, although in the case of swarms requires transforms that can handle time-varying graphs \cite{qiu2017time,bohannon2019filtering,ji2019hilbert}.  Multiscale transforms, such as graph wavelets, may be useful for identifying multiscale collective behaviors \cite{shuman2015multiscale,zheng2019framework}. Finally, we point out that signals of interest in a swarm may be functions of multiple agents (such as distance between agents) in which case extensions to \ac{GSP} that operate on edge signals \cite{schaub2018flow}, hypergraphs \cite{barbarossa2016introduction,zhang2019introducing}, and simplicial complices \cite{barbarossa2020topological} could be applied.  We note that simplicial complices are heavily involved in the underpinnings of \ac{TDA} and the swarm analysis approaches of \cite{topaz2015topological,sinhuber2017phase}.

\section*{Acknowledgments}
This work was supported by NSF award NCS/FO 1835279 and JHU/APL internal research and development funds. This material is based upon work supported by (while GH was serving at) the National Science Foundation. Any opinion, findings, and conclusions or recommendations expressed in this material are those of the author(s) and do not necessarily reflect the views of the National Science Foundation.

\bibliography{references.bib}

\clearpage
\onecolumngrid
\appendix

\section{Sample Graph Harmonics}
Below are higher-order graph harmonics for several different swarm states, in this case using swarmalator \cite{o2017oscillators} states.

\begin{figure}[!ht]

  \centering 
  \includegraphics[width=\columnwidth]{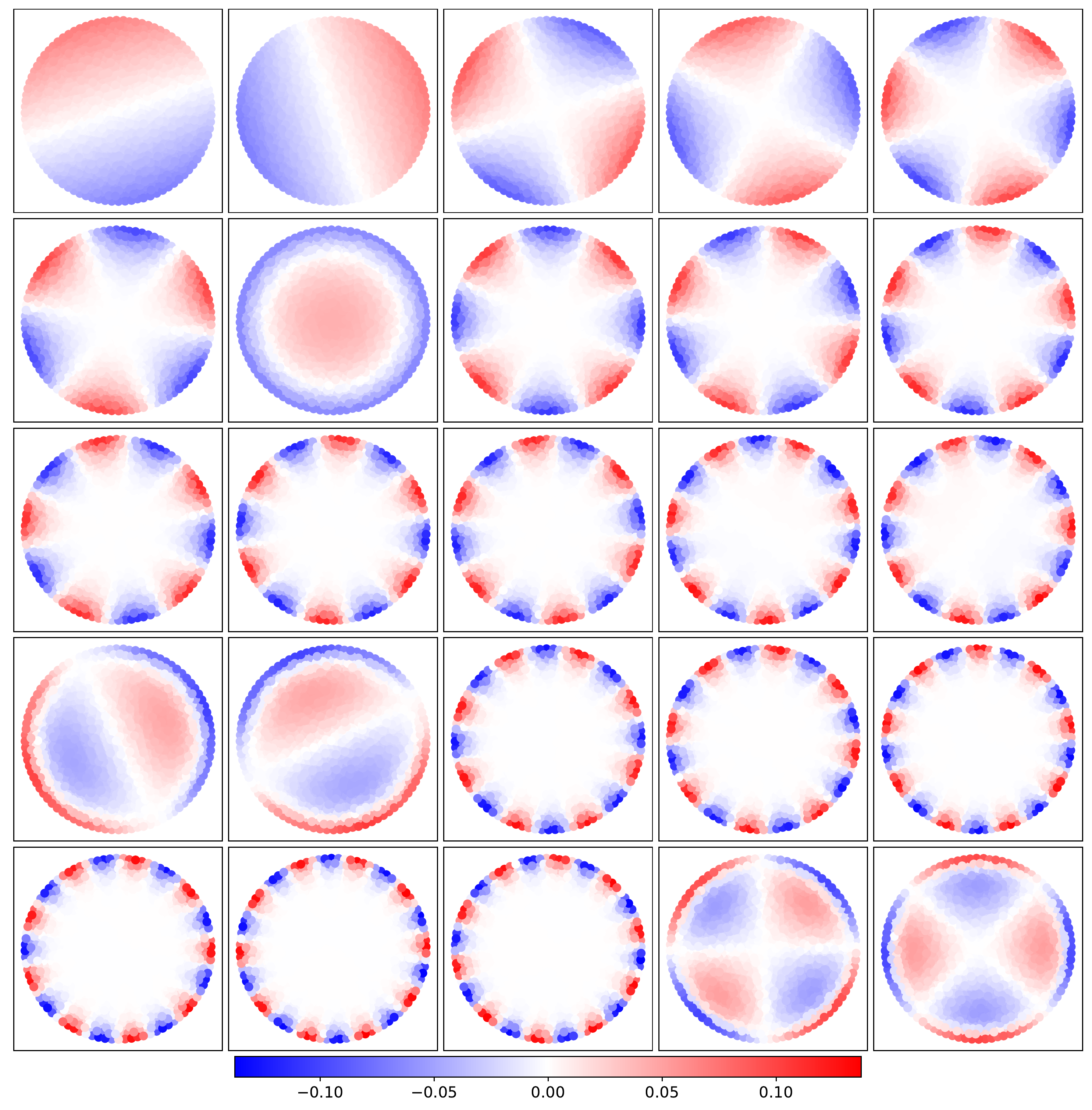}

  \caption{The first 25 harmonics of a disk state of 1000 swarmalator agents $A=B=K=1$, $J=0.1$, in increasing order from left to right, top to bottom.}

\end{figure}

\begin{figure}[!ht]

  \centering 
  \includegraphics[width=\columnwidth]{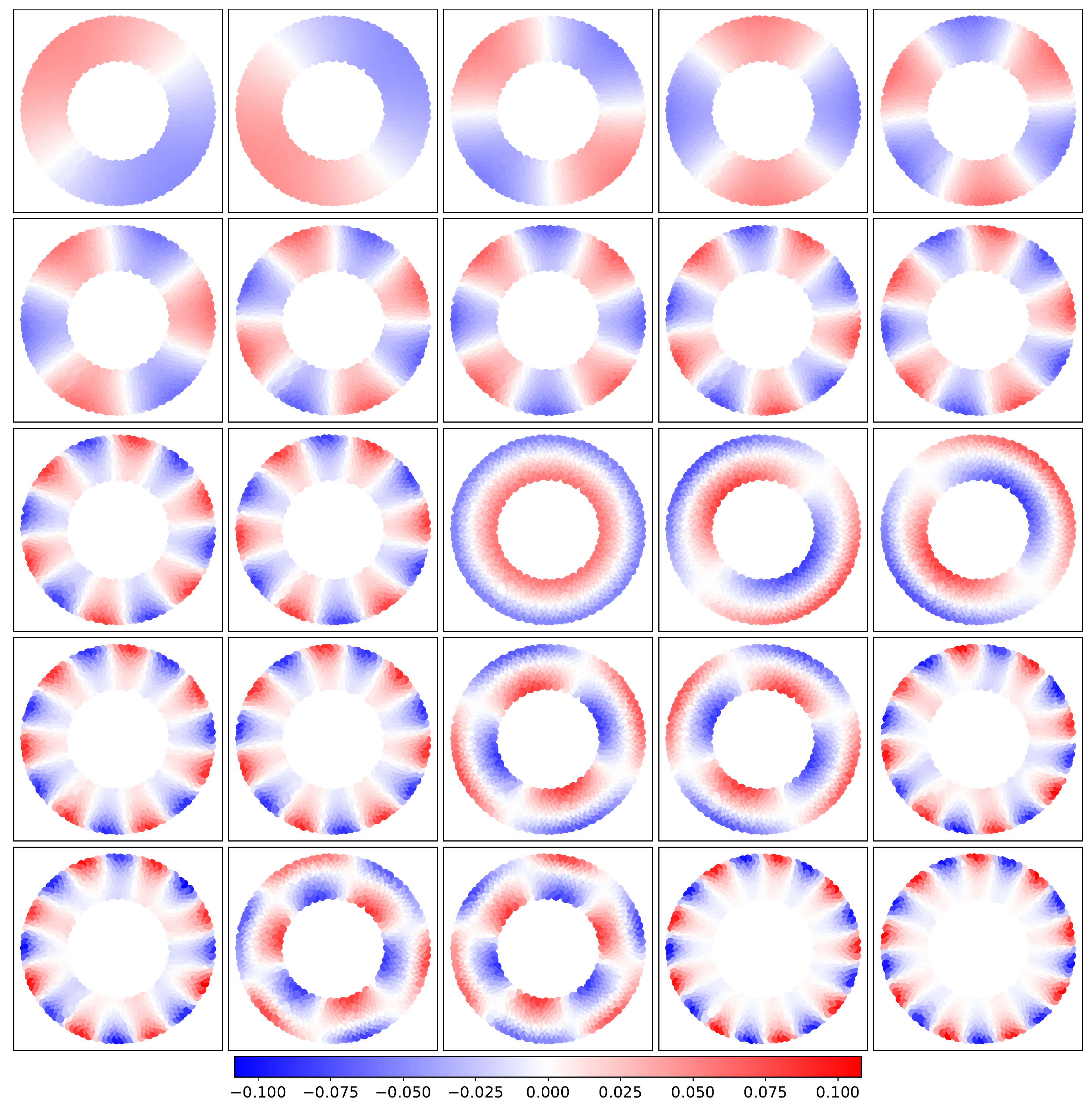}

  \caption{The first 25 harmonics of an annular state (standing phase wave state) of 1000 swarmalator agents $A=B=J=1$, $K=0$, in increasing order from left to right, top to bottom.}
\end{figure}

\begin{figure}[!ht]

  \centering 
  \includegraphics[width=\columnwidth]{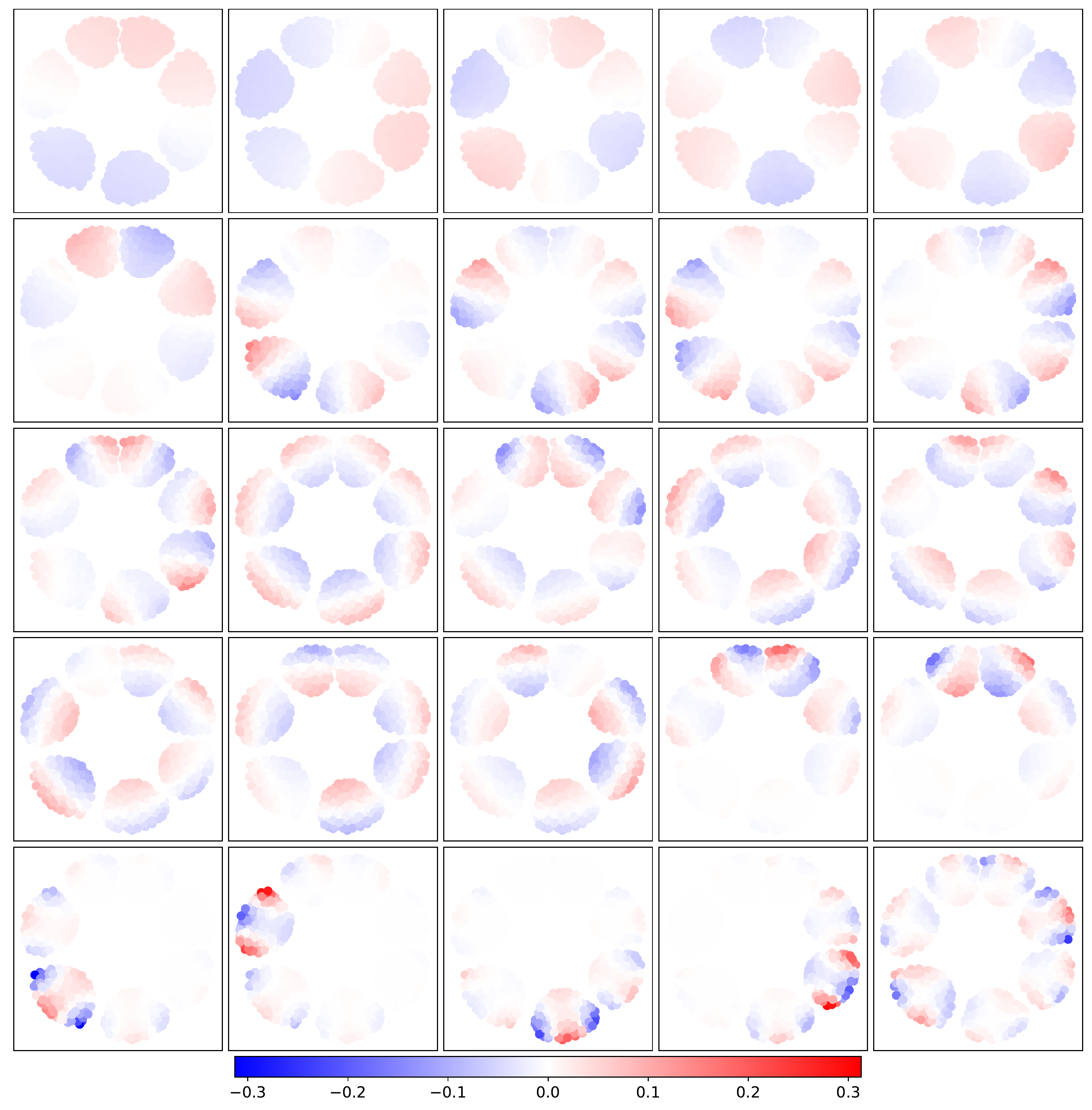}

  \caption{The first 25 harmonics of an annular state (standing phase wave state) of 1000 swarmalator agents $A=B=J=1$, $K=-0.1$, in increasing order from left to right, top to bottom.}
\end{figure}

\begin{figure}[!ht]

  \centering 
  \includegraphics[width=\columnwidth]{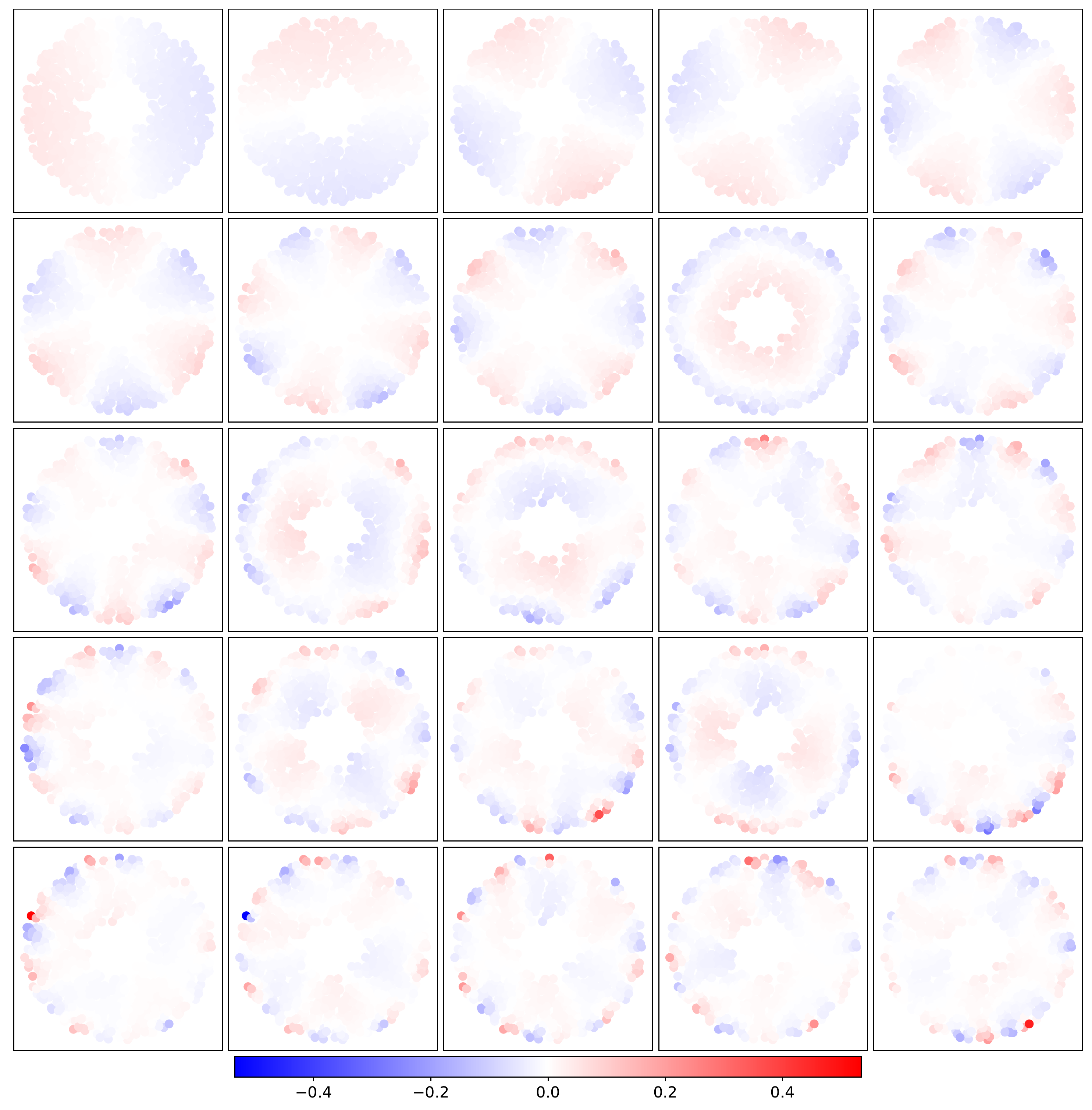}

  \caption{The first 25 harmonics of an annular state (standing phase wave state) of 1000 swarmalator agents $A=B=J=1$, $K=-0.75$, in increasing order from left to right, top to bottom.}
\end{figure}

\clearpage
\section{Solid Notional Torus}
This section contains a repeat of the analysis for a notional torus state that is filled, rather than hollow. 

\begin{figure}[!ht]

  \centering
  \includegraphics[width=\columnwidth]{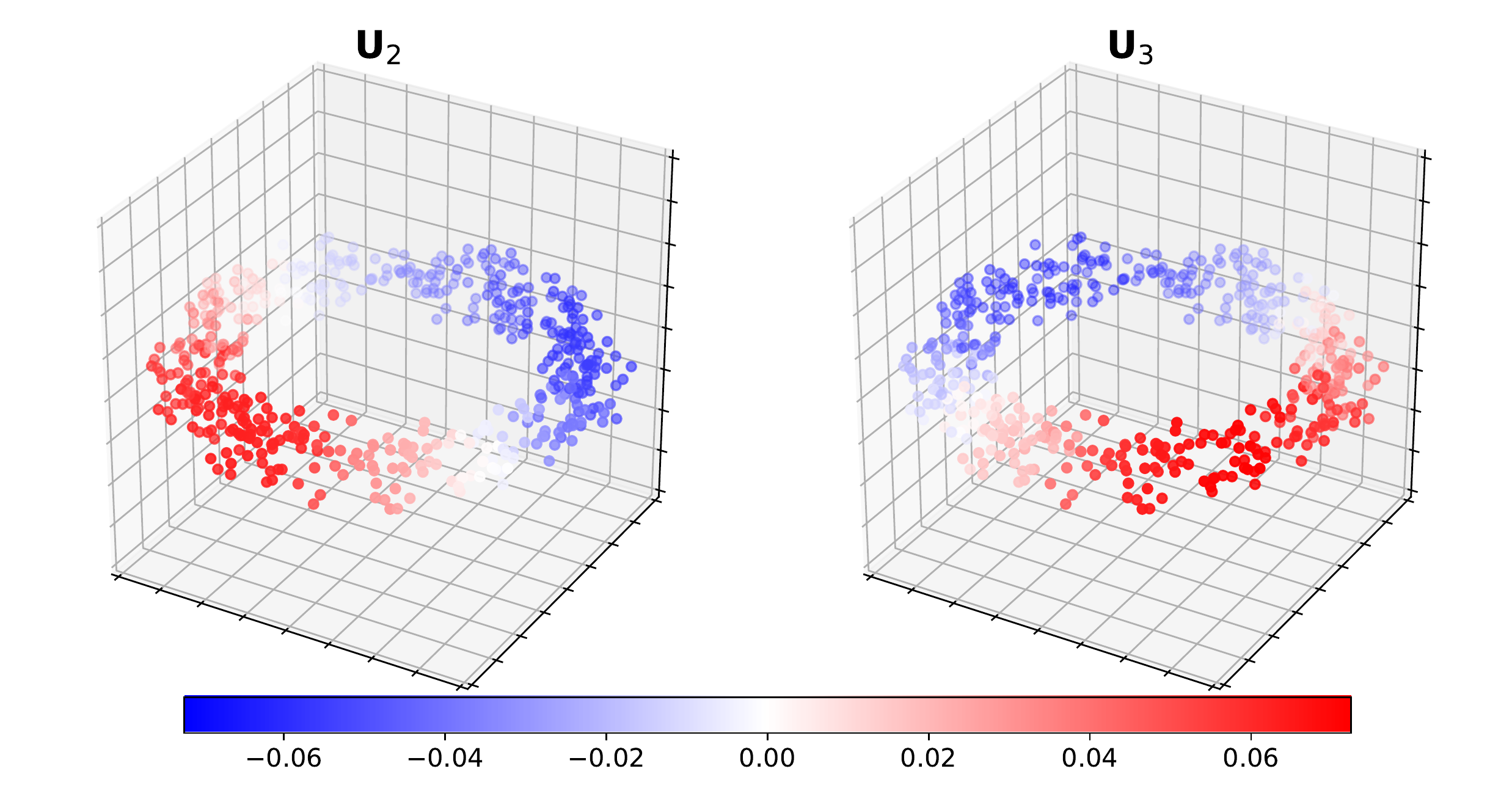}

  \caption{Second and third \ac{GFT} harmonics for a randomly generated notional solid torus state.$N=512$ agents were placed  uniformly at random inside a torus with toroidal radius one unit and poloidal radius $\frac{1}{4}$ units.  The adjacency matrix was defined by $A_{ij}=1$ if $||x_i-x_j||<\frac{1}{4}$ and 0 otherwise ($i\neq j$), $A_{ii}=0$. The combinatorial Laplacian $\mathbf{L}$ was used to define the \ac{GFT}. Note that these harmonics have period one with respect to the toroidal direction and are approximately $90^\circ$ out of phase.}

  \label{fig:torus_harmonics_solid}

\end{figure}

\begin{figure}[!ht]

  \centering
  \includegraphics[width=\columnwidth]{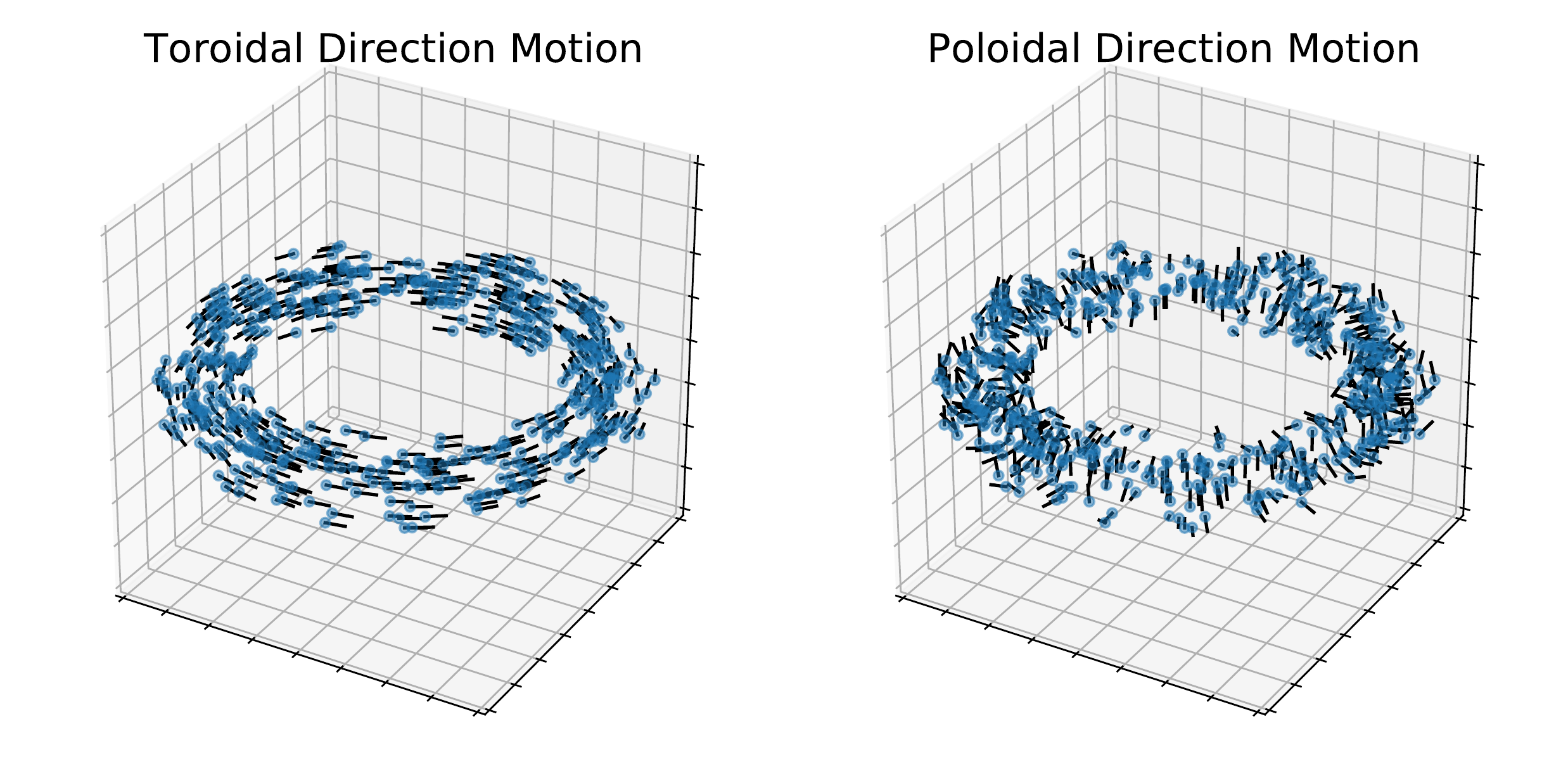}

  \caption{Example motion states along a solid torus. Left: Coherent motion in the toroidal direction. Right: Coherent motion in the poloidal direction.}

  \label{fig:torus_motion_solid}

\end{figure}

\begin{figure}[!ht]

  \centering
  \includegraphics[width=\columnwidth]{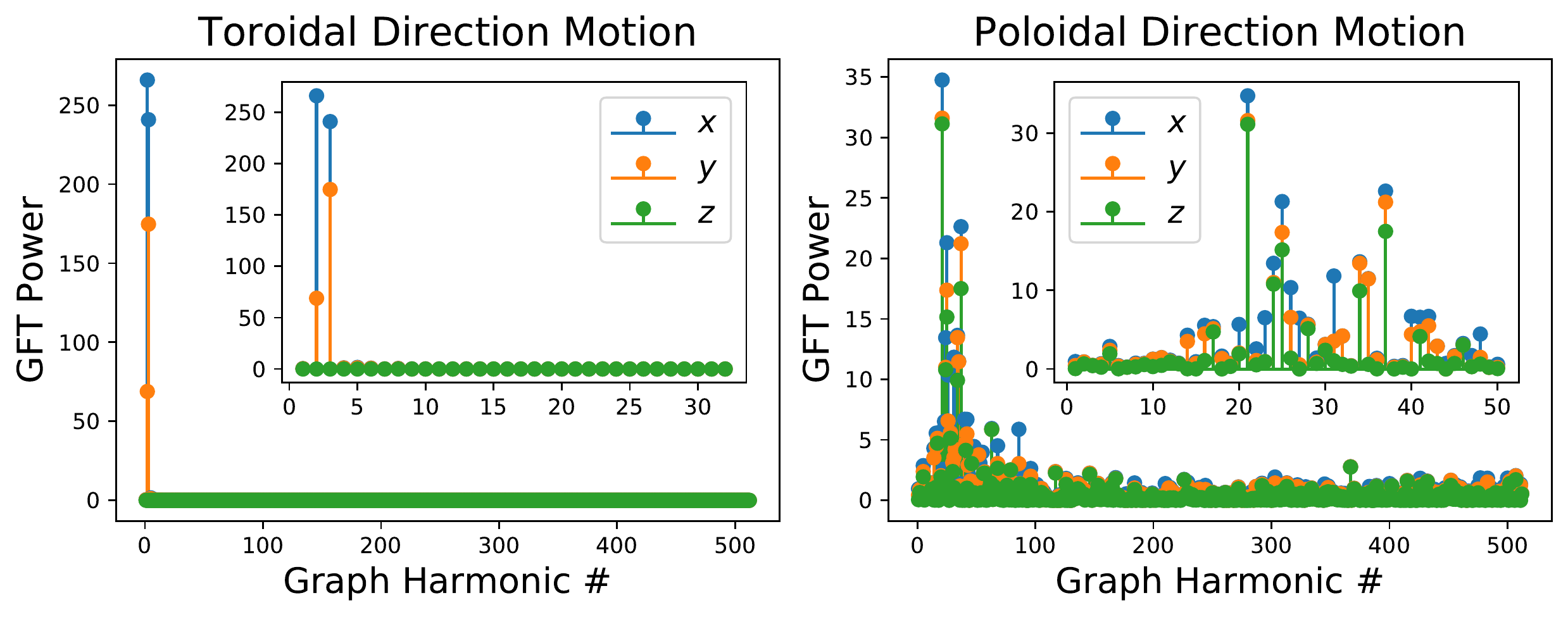}

  \caption{Stem plots of \ac{GFT} power for motion along toroidal direction (left) and poloidal direction (right). \ac{GFT} power is shown decomposed into the $x$, $y$, and $z$ dimensions, with the height of each color indicating the contribution from that dimension in a stacked fashion.  Since the toroidal direction motion has no $z$ components, its contribution to the \ac{GFT} power is zero, and the \ac{GFT} structure is essentially the same as in the annular states.  The poloidal direction is more complex and while still a ``low-frequency'' signal, has higher frequency contributions from all three axes (notably, their are negligible contributions in the first through third harmonics).}

  \label{fig:torus_GFT_solid}

\end{figure}

\begin{figure}[!ht]

  \centering
  \includegraphics[width=.8\columnwidth]{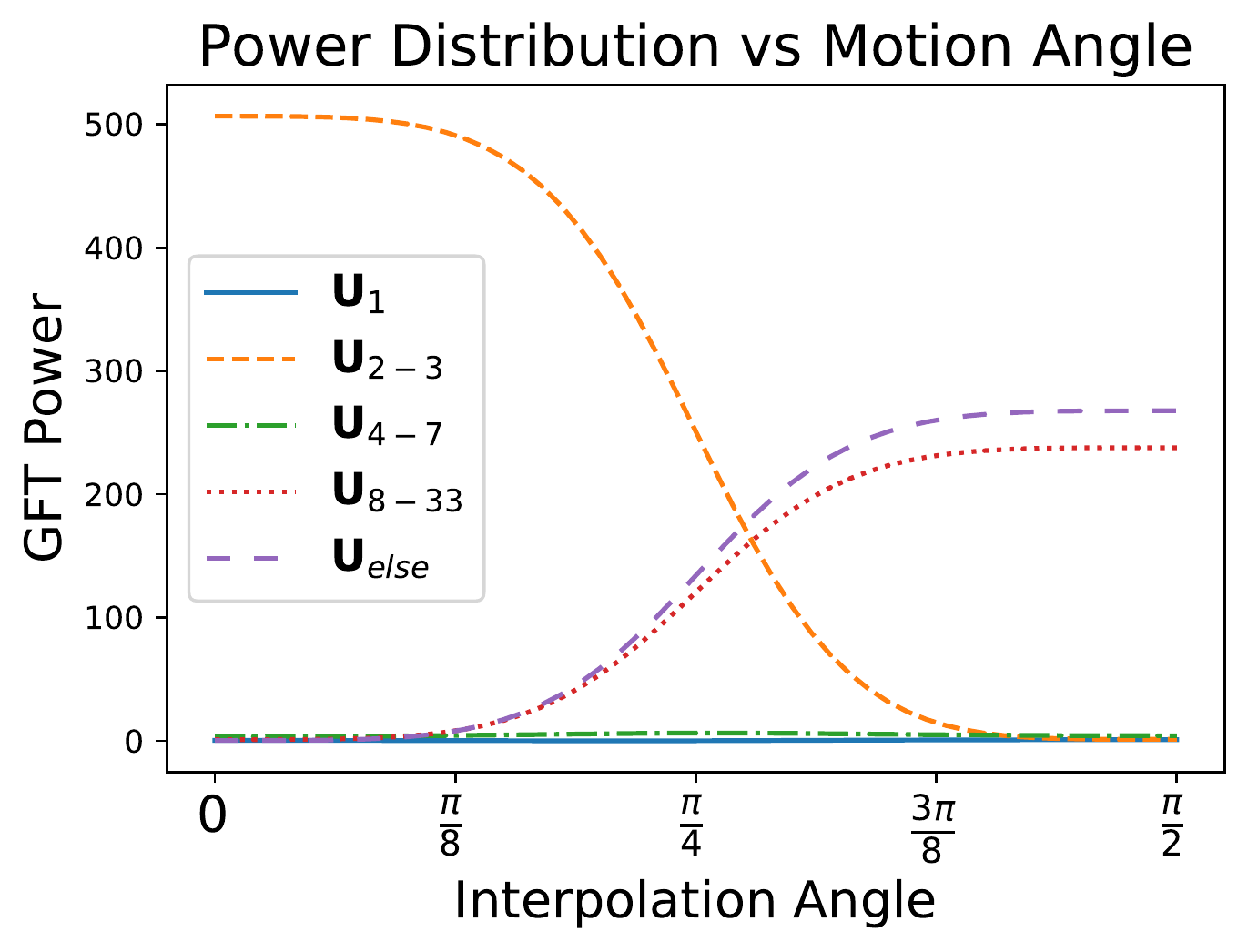}

  \caption{Transition of power in \ac{GFT} harmonics as the motion along the notional torus shape is interpolated from motional along the toroidal direction to motion along the poloidal direction. Since the \ac{GFT} is linear, a linear combination of the two perpendicular directions results in a linear combination in the \ac{GFT} power.}

  \label{fig:torus_interp_solid}

\end{figure}

\clearpage
\section{Curve States}

\begin{figure}[!ht]

  \centering
  \includegraphics[width=\columnwidth]{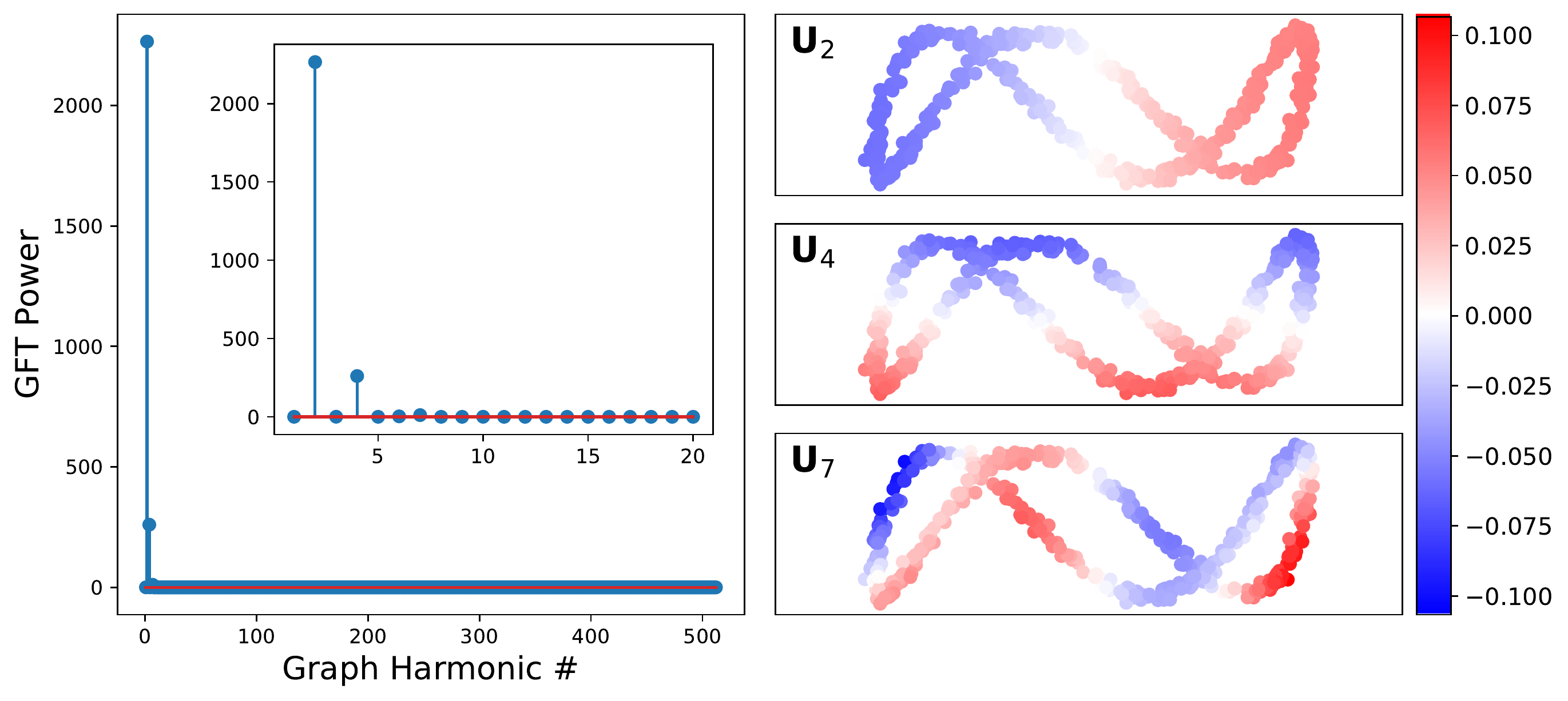}

  \caption{\ac{GFT} power of $\mathbf{r}$ and top three harmonics derived using the normalized Laplacian of a squared-exponentially weighted graph.}

  \label{fig:curve_state_GFT_harms_r}

\end{figure}

\clearpage
\section{Vicsek-like Model}
The \ac{GFT} power of the five milling state of Example~1 using a squared exponential weighting (as opposed to the sensing range) produces similar spectral signatures but it is less clear what constitutes a connected component in this context.  The harmonics themselves appear to be qualitatively similar to that in Example~1.   This illustrates that the analysis is at least somewhat robust to the particular choice of graph.

\begin{figure}[!ht]

  \centering
  \includegraphics[width=.6\columnwidth]{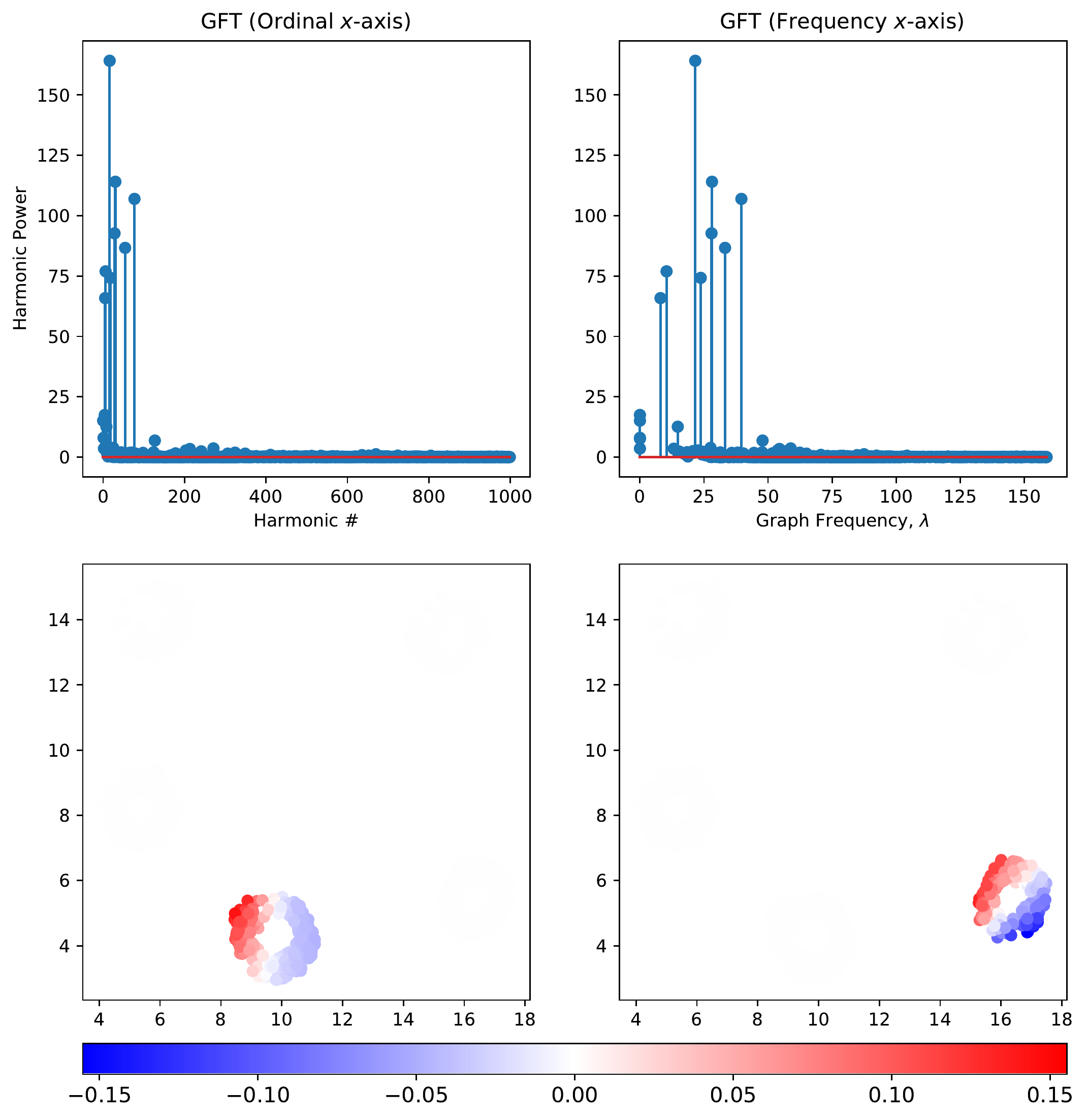}

  \caption{Graph Fourier Analysis using the same state as Fig.~\ref{fig:vicsek_five} but using edge weighs $A_{ij}=\exp(-||\mathbf{x}_i-\mathbf{x}_j||^2_2)$ instead of the hard threshold. Top: Harmonic power for each harmonic displayed ordinally by graph frequency (i.e., $\lambda_i$) and by frequency. Bottom: Sample graph harmonics for corresponding to the strongest (left) and fourth strongest (right) power.  These demonstrate that the analysis is at least somewhat stable for different choices of $\mathbf{A}$, but also highlight the numerical challenges present in dealing with swarms that are essentially disconnected.}

\end{figure}

 \clearpage
 \section{Couzin Model}

Supporting figures for Example 2.
 
\begin{figure}[!ht]
 
 \centering 
   \includegraphics[width=\columnwidth]{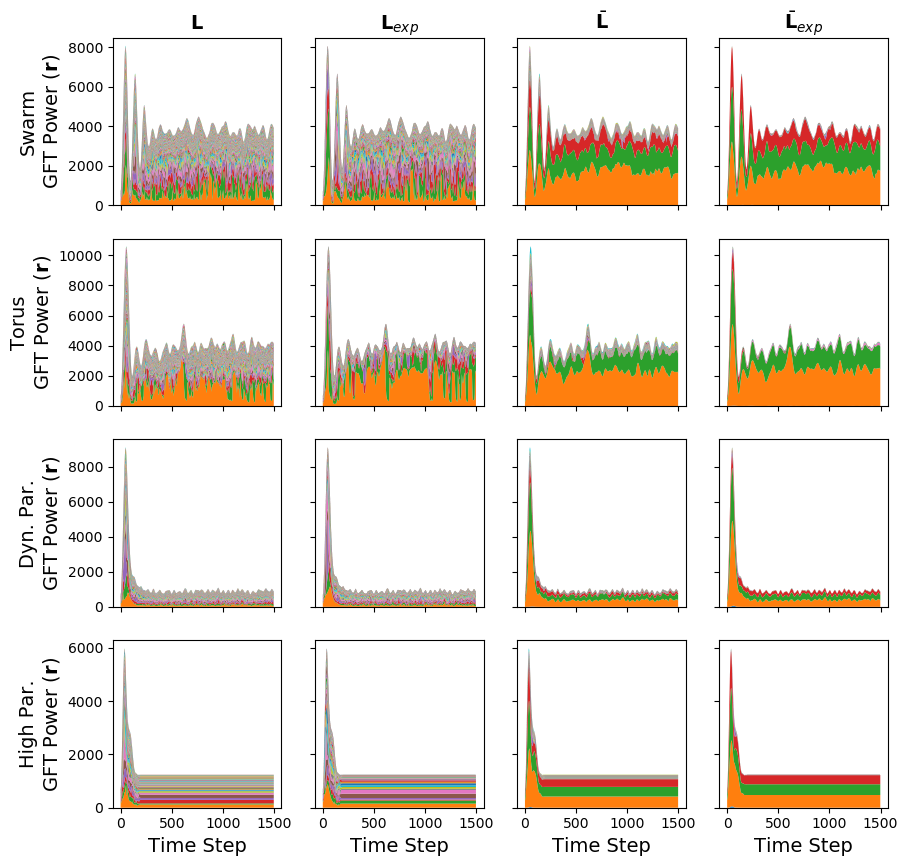}
 
 \caption{Stacked area plots of the \ac{GFT} power (using four different transforms (columns)) of $\mathbf{r}$ of simulations of \cite{couzin2002collective} for a simulation run of each of the four collective behaviors (Swarm, Torus, Dynamic Parallel, and Highly Parallel (rows)). Note there is no power in the first harmonic so the large orange, green, and red regions correspond to the second, third, and fourth harmonics, respectively. Note that the normalized versions are more stable over time than the others, and have more power in harmonics 2-4 than the non-normalized versions.  Again, the results appear relatively insensitive to the particular choice of graph.}
 
 \end{figure}
 
 \begin{figure}[!ht]
 
   \centering 
   \includegraphics[width=\columnwidth]{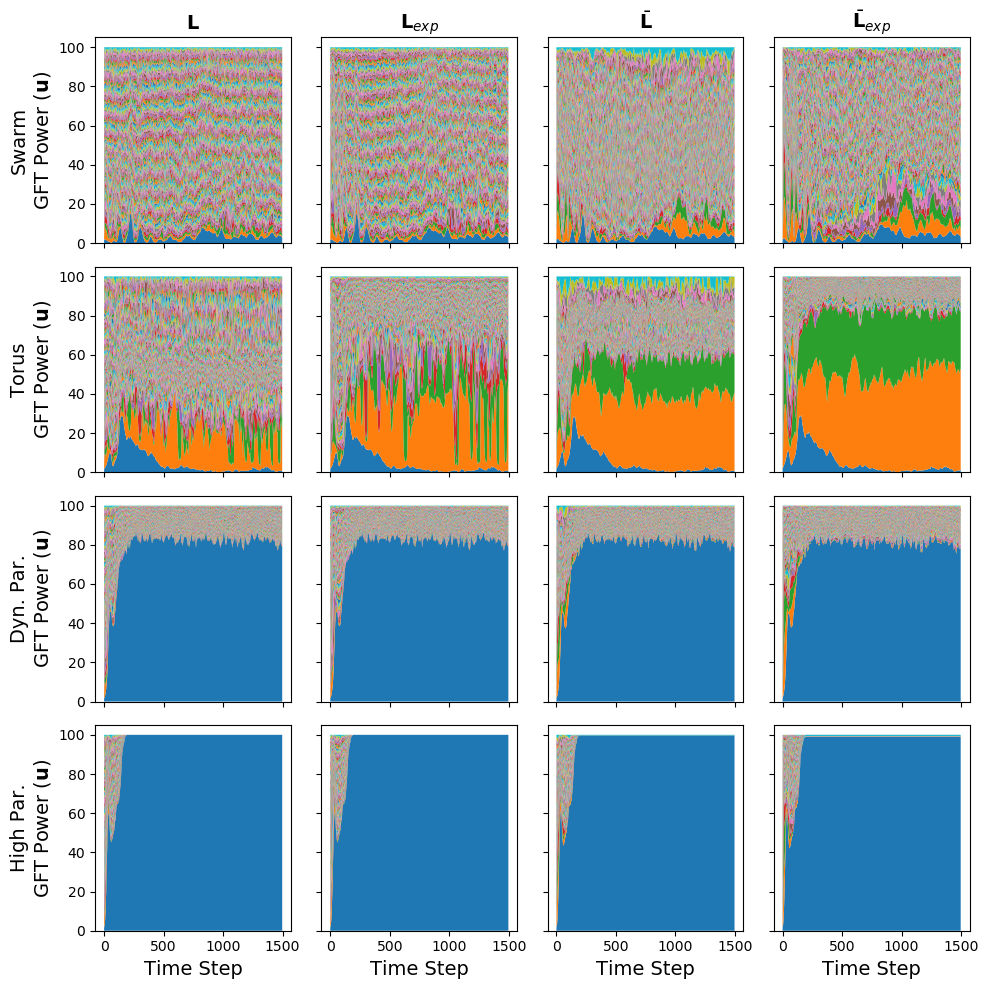}
 
  \caption{Stacked area plots of the \ac{GFT} power (using four different transforms (columns)) of $\mathbf{u}$ of simulations of \cite{couzin2002collective} for a simulation run of each of the four collective behaviors (Swarm, Torus, Dynamic Parallel, and Highly Parallel (rows)). Note there is no power in the first harmonic so the large orange, green, and red regions correspond to the second, third, and fourth harmonics, respectively. Note that the normalized versions are more stable over time than the others, and have more power in harmonics two and three than the non-normalized versions for the Torus state.  Again, the results appear relatively insensitive to the particular choice of graph.}
 \end{figure}
\end{document}